\documentclass[%
 reprint,
 superscriptaddress,
 amsmath,amssymb,
 aps,
 floatfix,
]{revtex4-2}

\usepackage{xcolor}

\newcommand{\xox}[1]{\sigma_x {#1} \sigma_x}
\newcommand{\yoy}[1]{\sigma_y {#1} \sigma_y}
\newcommand{\zoz}[1]{\sigma_z {#1} \sigma_z}

\newcommand{\sx}{\sigma_x}
\newcommand{\sy}{\sigma_y}
\newcommand{\sz}{\sigma_z}

\usepackage{float}

\usepackage{tikz}
  \usetikzlibrary{positioning}
  \usetikzlibrary{quantikz}

\usepackage{amsmath}
\usepackage{mathtools}

\tikzset{
operator/.append style={rounded corners},
}

\usepackage{graphicx}
\usepackage{dcolumn}
\usepackage{bm}
\usepackage{physics}
\usepackage{dsfont}

\usepackage{amsmath}

\makeatletter
\newsavebox{\@brx}
\newcommand{\llangle}[1][]{\savebox{\@brx}{\(\m@th{#1\langle}\)}%
  \mathopen{\copy\@brx\kern-0.5\wd\@brx\usebox{\@brx}}}
\newcommand{\rrangle}[1][]{\savebox{\@brx}{\(\m@th{#1\rangle}\)}%
  \mathclose{\copy\@brx\kern-0.5\wd\@brx\usebox{\@brx}}}
\makeatother

\newtheorem{corollary}{Corollary}

\begin{document}

\title{Qubit noise deconvolution}

\author{Stefano Mangini}
\email{stefano.mangini01@universitadipavia.it}
\affiliation{Dipartimento di Fisica, Università di Pavia, Via Bassi 6, I-27100, Pavia, Italy}
\affiliation{INFN Sezione di Pavia, Via Bassi 6, I-27100, Pavia, Italy}

\author{Lorenzo Maccone}
\affiliation{Dipartimento di Fisica, Università di Pavia, Via Bassi 6, I-27100, Pavia, Italy}
\affiliation{INFN Sezione di Pavia, Via Bassi 6, I-27100, Pavia, Italy}
\affiliation{CNR-INO -  Largo E. Fermi 6, I-50125, Firenze, Italy}

\author{Chiara Macchiavello}
\affiliation{Dipartimento di Fisica, Università di Pavia, Via Bassi 6, I-27100, Pavia, Italy}
\affiliation{INFN Sezione di Pavia, Via Bassi 6, I-27100, Pavia, Italy}
\affiliation{CNR-INO -  Largo E. Fermi 6, I-50125, Firenze, Italy}

\date{\today}

\begin{abstract}
We present a noise deconvolution technique to remove a wide class of noises when performing arbitrary measurements on qubit systems. In particular, we derive the inverse map of the most common single qubit noisy channels, and exploit it at the data processing step to obtain noise-free estimates of observables evaluated on a qubit system subject to known noise. We illustrate a self-consistency check to ensure that the noise characterization is accurate providing simulation results for the deconvolution of a generic Pauli channel, as well as experimental evidence of the deconvolution of decoherence noise occurring on Rigetti quantum hardware. 
\end{abstract}

\maketitle

\section{\label{sec:Intro} Introduction}
Quantum noise is currently the largest limiting factor in the adoption of quantum computation and quantum technology. Their theoretical performances are in fact hindered by the intrinsic fragility of quantum systems, and over the last years many proposal have been put forward to mitigate, ideally correct, the effect of noise and recover reliable results. On the computing side, as fault-tolerant quantum computers remain out of reach at the moment~\cite{FT_EC_Steane, FT_EC_Shor, knillQuantumComputingRealistically2005, PreskillNISQ}, various error mitigation techniques have been proposed to extend the capabilities of current small scale noisy quantum devices~\cite{suzuki2021quantum, Cao_NISQMitigation, DynamicalDecoupling}. These ranges from correcting the readout noise via inversion of probability assignment matrix~\cite{ReadoutErrorMitigation}, extrapolating the noise in the device to the zero error case~\cite{ErrorMitigationMari, Temme_PEC, ZNE_QEM}, using a probabilistic sampling on specific circuits to approximate the noise free computation~\cite{Temme_PEC, Endo_Mitigation, ErrorMitigationMari}, to also using machine learning approaches to learn how to recover ideal results~\cite{ML_QEM}.

While these methods are concerned with mitigating noise occurring in a computation, here we instead focus on the more generic task of correcting the expectation value of arbitrary observables evaluated on a system which is subject to a known noise happening before the measurement stage. Such a scenario is relevant in quantum communication and quantum tomography tasks~\cite{Vikesh_MitigationTomography}. 

Noise in quantum systems is described by means of quantum channels~\cite{NielsenChuang}
\begin{equation}
\label{eq:Kraus}
    \rho \longrightarrow \mathcal{E}(\rho) = \sum_k A_k \rho A_k^\dagger\,,
\end{equation}
where $A_k$ are operators acting on the system named Kraus operators.
While the effect of unitary dynamics can be reversed using realizable operations, quantum channels cannot be undone, and one can only hope to find operations which only approximately invert the noise process at hand. Examples of this approach leverages for example Petz recovery maps~\cite{PetzMaps_Inversion, gilyen2020quantum, Wilde_QIT}, or unitaries which, on average, are able to best reverse the noise based on given distance measures~\cite{Benatti_Inversion, Zyczkowski_Inversion, Aurell_2015}.

\begin{figure}[htbp]
    \centering
\begin{tikzpicture}
\node at (-3,0.5) {\textbf{(a)}};
\node at (0,0) (circ) {\begin{quantikz}
     \lstick{$\rho$} & \gate{M}\gategroup[1,steps=2,style={dashed,
                   rounded corners,fill=blue!5, inner xsep=0pt},
                   background]{$O$} &[-0.25cm] \meter{} & \cw \rstick{$\Tr[O\rho]$}
\end{quantikz}};
\node at (4, -0.25) {Ideal};
\node at (-3,-1.5) {\textbf{(b)}};
\node at (0,-2) {\begin{quantikz}
     \lstick{$\rho$} & \gate[style={fill=yellow!20}]{\mathcal{N}}  &\gate{M}\gategroup[1,steps=2,style={dashed,
                   rounded corners,fill=blue!5, inner xsep=0pt},
                   background]{$O$} &[-0.25cm] \meter{} & \cw \rstick{$\Tr[O\mathcal{N}(\rho)]$}
\end{quantikz}};
\node at (4, -2.25) {Noisy};

\node at (-3,-3.5) {\textbf{(c)}};
\node at (0,-4) {\begin{quantikz}
     \lstick{$\rho$} &  \gate[style={fill=yellow!20}]{\mathcal{N}} & \gate{M}\gategroup[1,steps=2,style={dashed,
                   rounded corners,fill=blue!5, inner xsep=0pt},
                   background]{$\mathcal{N}^{-1}(O)$} &[-0.25cm] \meter{} & \cw \rstick{$\expval{O}$}
\end{quantikz}};
\node at (4, -4.25) {Deconvolution};

\node at (-3,-5.5) {\textbf{(d)}};
\node at (0, -6) {\begin{quantikz}
     \lstick{$\rho$} & \gate[style={fill=yellow!20}]{\mathcal{N}_0} & \gate{U} & \gate[style={fill=yellow!20}]{\mathcal{N}_1} & \meter{$O$} & \cw
\end{quantikz}};
\node at (4, -6.) {Only $\mathcal{N}^{-1}_1$};
\end{tikzpicture}
    \caption{General scheme for the noise deconvolution process applied to a qubit. \textbf{(a)} Ideal estimation of an observable $O$ on a single qubit in state $\rho$. The operator $M \in \{\mathds{1}, H, H S^\dagger\}$, with $H$ and $S$ being the Hadamard and Phase gate are used to select a measurement basis in $\{\sigma_z, \sigma_x, \sigma_y\}$ respectively, and thus reconstruct a generic observable $O$, using Eq.~\eqref{eq:qubit_tomography}. \textbf{(b)} Noise (indicated with a yellow box) happening before measurement leads to noisy estimates of the expectation values. \textbf{(c)} Noise deconvolution approach: measurements of the noise-inverted observables $\mathcal{N}^{-1}(O)$ on the noisy state leads to the mitigated ideal result $\expval{O}$. \textbf{(d)} The noise deconvolution approach can be used to mitigate the effects of $\mathcal{N}_1$ only. However, the full noise ($\mathcal{N}_0$ and $\mathcal{N}_1$) can be mitigated either if the unitary can be easily inverted as well, or if the noise processes commutes with the interleaving unitary, as is the case for depolarizing noise. }
    \label{fig:noise_deconvolution_summary}
\end{figure}
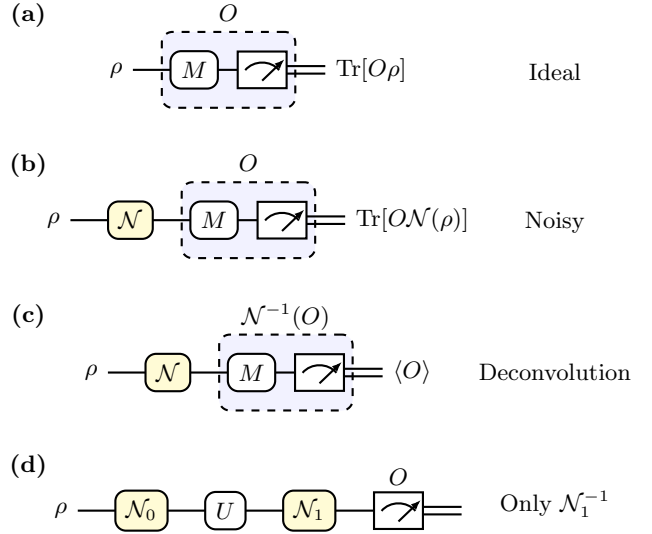

Here instead we show that noise can be eliminated by means of a \textit{deconvolution} process, provided that the noise map describing the process is known and invertible. In fact, we drop the requirements of the inverse transformation being itself a quantum channel, since the transformation is not applied to the quantum system itself,  but to the outcome statistics as a classical post-processing step. We derive the inversion maps of the most common single-qubit noisy channels (both unital and non-unital), and show how to use these to remove the effect of noise from the expectation values of general observables.  In Figure~\ref{fig:noise_deconvolution_summary} we schematically summarize the noise deconvolution idea. The mitigation is effectively obtained by multiplying the noisy estimates by a factor depending on the noise $\expval{O}_{mitig} \sim c \expval{O}_{noisy}$, which comes at the cost of increasing the variance of the estimation, as $\text{Var}[\expval{O}_{mitig}] \sim c^2 \text{Var}[\expval{O}_{noisy}]$, so one needs to gather more statistic to reach a target precision. A related post processing technique specialized for quantum many-body systems and quantum field theory is put forward in ref.~\cite{benyQuantumDeconvolution2017}. In addition, we provide both numerical simulations of the noise deconvolution process, as well as evidence of deconvolution of decoherence noise occurring on the superconducting quantum computer ``Aspen-9'' provided by Rigetti, accessed using the Quantum Cloud Services (QCS)~\cite{Karalekas_2020}. We show how simple self-consistency checks can test whether the known noise map is accurate and how a feedback scheme can be used to adjust the noise parameters.
Our contributions then include: (\textit{i}) formalization and discussion of CPTP noise deconvolution of expectation values through (mathematical) inversion of the noise map; (\textit{ii}) explicit derivation of the inverse map of the most common single qubit noise channels; (\textit{iii}) numerical and experimental application of the ideas introduced before. 

Before continuing, we briefly describe the the relation of the proposed noise deconvolution idea to probabilistic error mitigation (PEC)~\cite{Temme_PEC, Endo_Mitigation}, a quantum error mitigation technique aimed at correcting noisy operations during a quantum computation. Given a characterization of the noise, PEC works by using the inverse noise map of the operations to build an ensemble of suitably generated quantum circuits. These are sampled according to specific weights, and the results combined to build an approximation of the action of the noise-free quantum circuit. In particular, the mitigation procedure is \textit{active}, in the sense that the experimenter need to generate new quantum circuits and run them against the quantum device. 
On the contrary, we are instead concerned with the correction of expectation values evaluated on a noisy state, with no computation or dynamics involved. In addition, within our framework, the mitigation is \textit{passive}, in the sense that the mitigation happens classically as a post-processing step, and no action on the quantum system is necessary. Appropriately limiting PEC to the specific case of measurement error mitigation, and realizing that sampling on quantum circuits is no longer a necessary step, then one can recover the noise deconvolution procedure presented here, whose regime of application is not restricted to quantum computation, but applies to a general quantum mechanical measurement scenario. As such, some of the results presented here can be recovered also with the techniques proposed in~\cite{Temme_PEC, Endo_Mitigation}. That said, the explicit calculations presented here for the general noise maps we analyze have not been presented elsewhere in full generality, e.g. see Table~\ref{tab:tablesummary} below.

The Letter is organized as follows. In Sec.~\ref{sec:Preliminaries} we recall some basic concepts about quantum channels and the Pauli transfer matrix formalism, and the idea of noise deconvolution in Sec.~\ref{sec:NoiseDeconvolution}. In Sec.~\ref{sec:MapInversion} we leverage the Pauli transfer matrix formalism to explicitly derive the inverse map of the most common single qubit noise channels, and use inside the noise deconvolution procedure to obtain noise-free estimates. In Table~\ref{tab:tablesummary} we summarize all the maps taken in consideration as well as their inverse. In Sec.~\ref{sec:Experiments} we show by means of simulations that the noise deconvolution process can be used to cancel out the effect of a general Pauli channel, and also provide experimental evidence of the deconvolution of decoherence noise as performed on a real quantum device by Rigetti. 

\section{\label{sec:Preliminaries} Methods}
In this section we introduce the notation and the theoretical tools used to derive the main results of the work. We will denote with $\mathcal{H}$ the Hilbert space, and with $\mathcal{L}(\mathcal{H})$ the space of squared linear operators acting on $\mathcal{H}$. For those interested, a brief overview of quantum channels and Kraus decomposition can be found in Appendix~\ref{app:KarusDecomposition}.

\subsection{\label{subsec:Channels} Quantum channels}
In general quantum channels cannot be physically inverted, as there is no quantum evolution capable of reversing their actions. Formally stated, let $\mathcal{E}$ be a CPTP map, it is not possible to find another CPTP map $\mathcal{D}=\mathcal{E}^{-1}$, such that $(\mathcal{D \circ \mathcal{E}})(\rho) = \rho \, \forall\, \rho$. The only trivial case when this is possible, is for maps having only a single Kraus operator, in which case they reduce to standard unitary evolution $\mathcal{E}(\rho) = U\rho U^\dagger$, with the inverse given by $\mathcal{D}(\cdot) = U^\dagger\,(\cdot)\,U$.

The CPTP conditions impose hard constraints to the operatorial form that physically realizable evolutions must match, namely the Kraus representation. However, the requirement for admitting a more general operator-sum representation are looser. In fact, any Hermiticity preserving map, i.e. a map such that $\Phi(\rho)^\dagger = \Phi(\rho)$ for $\rho=\rho^\dagger$, admits an operator-sum representation as~\cite{JohndepillisLinearTransformationsWhich1967, BourdonUnitalQuantumOperations2004} 
\begin{equation}
\label{eq:general_map}
    \Phi(\rho) = \sum_{k} \lambda_k A_k \rho A_k^\dagger
\end{equation}
with $\lambda_k \in \{+1,-1\}$. Clearly if all the coefficients are $\lambda_k = 1 \, \forall k$, then the map $\Phi$ is also completely positive, since it is in standard Kraus form~\eqref{eq:Kraus}. Moreover, another useful characterization is given by
\begin{corollary}[Corollary II.2 of ~\cite{BourdonUnitalQuantumOperations2004}]
Let $\mathcal{M}_N$ be the space of complex $N\times N$ matrices. Suppose $\Phi:\mathcal{M}_N\rightarrow \mathcal{M}_N$ is a completely positive map having the form
\begin{equation}
\label{eq:corollary}
    \Phi(\rho) = \sum_{k}\beta_k A_k \rho A_k ^\dagger
\end{equation}
where $\{A_k\}_{k}$ is linearly independent in $\mathcal{M}_N$, and $\beta_k \in \mathbb{R} \, \forall k$. Then $\beta_k \geq 0 \, \forall k$.
\end{corollary}

Conversely, if a map has form~\eqref{eq:corollary} with linearly independent operators $\{A_k\}_k$ but has \textit{some} of the coefficients $\beta_j<0$, then the map is not completely positive. This result is steadily applied to maps acting on qubit systems where $\mathcal{M}_N = \mathbb{C}^{2\times2}$. In fact, Pauli matrices $\sigma_x, \sigma_y$ and $\sigma_z$ together with the identity $\sigma_0 = \mathds{1}_2$ form a linearly independent set in the space of $2\times2$ complex matrices, and then any map of the form 
\begin{equation}
\label{eq:general_NON_CPTP_map}
    \mathcal{E}(\rho) = \beta_0 \sigma_0O\sigma_0 + \beta_1 \xox{\rho} + \beta_2 \yoy{\rho} + \beta_3 \zoz{\rho}\, 
\end{equation}
having some negative coefficients is not a CP map, thus it is not a physically realizable channel. In the following we will derive many inverse maps having this form, for which this result holds. Of course, we already know that a quantum channel cannot be inverted (apart from the trivial unitary case), so that if an inversion map is found, then  it is certainly not CP. Nonetheless, this result is still of interest because it gives a nice and clear condition that can be used to quickly assess the nature of the maps under investigation. In addition, as shown in ref.~\cite{Jiang2021_QuantumMaps}, if a CPTP map is invertible, then its inverse is Hermitian preserving (HP), and so can be expressed in operator-form~\eqref{eq:general_map}.

\subsection{Qubit systems and Pauli Transfer Matrix formalism} 
We focus on quantum systems made of qubits, and briefly review some useful results on qubit channels.  
The identity and the Pauli matrices $\{\mathds{1}, \sigma_x, \sigma_y, \sigma_z\}$ form a basis on $\mathcal{L}(\mathcal{H})=\mathbb{C}^{2\times 2}$, so that any operator $O \in \mathcal{L}(\mathcal{H})$ can be expressed in this basis as $O = o_0 \mathds{1} + \bm{o}\cdot{\bm{\sigma}}$, with $\bm{\sigma} = (\sigma_x, \sigma_y, \sigma_z)$, and $\bm{o} \in \mathbb{C}^3$. Similarly, density operators are expressed in this basis in terms of their Bloch vector as $\rho = \frac{\mathds{1}+\bm{r}\cdot\bm{\sigma}}{2}$, with $\bm{r}=(r_x, r_y, r_z) \in \mathbb{R}^3$, and $|\bm{r}|\leq 1$, where equality holds only for pure states $\rho = \dyad{\psi}$. Thus, in this basis, any operator $O$ (so both observables and states)  can be regarded as a column vector $|O\rrangle = [o_0, o_1, o_2, o_3]^\mathsf{T}$ whose components are given by $o_i = \Tr[\sigma_i\, O],\, i=0,1,2,3$ with $\sigma_0 = \mathds{1}$.

In addition, every linear map $\Phi: \mathcal{L}(\mathbb{C}^2) \rightarrow \mathcal{L}(\mathbb{C}^2)$ can be represented in this basis as a $4\times4$ matrix $\Gamma$~\cite{BourdonUnitalQuantumOperations2004,RuskaiAnalysisCompletelypositiveTracepreserving2002,KingMinimalEntropyStates2001, GreenbaumQuantumGate2015}, whose action is given by: 
\begin{equation}
\begin{aligned}
\label{eq:matrix_channel}
    \Phi(O) \rightarrow \Gamma |O\rrangle & = 
    \begin{bmatrix}
    \gamma_{0} & \bm{\gamma} \\
    \bm{t} & T \\
    \end{bmatrix}
    \begin{bmatrix}
    o_0 \\
    \bm{o}
    \end{bmatrix} \\
    & = \begin{bmatrix}
    \gamma_0 o_0 + \bm{\gamma}\bm{o}\\
    o_0\bm{t} + T\bm{o}
    \end{bmatrix} \\
    \qq{or} \Phi(O) & = (\gamma_0 o_0 + \bm{\gamma}\bm{o})\mathds{1} + (o_0\bm{t}+T\bm{o})\cdot \bm{\sigma}\, .
\end{aligned}
\end{equation}
Here $\bm{\gamma}$ and $\bm{t}$ being row and column vectors respectively, and $T$ is a $3\times3$ matrix. The $\Gamma$ matrix associated to the map $\Phi$ is called Pauli Transfer Matrix (PTM), and its elements are given by
\begin{equation}
\label{eq:PTM}
    \Gamma_{ij} = \frac{1}{2}\Tr[\sigma_i\Phi(\sigma_j)] \quad i,j \in \{0,1,2,3\}\, , \sigma_0 = \mathds{1}\,.
\end{equation}

If we restrict to trace-preserving maps, then $\bm{\gamma} = \bm{0}$ and $\gamma_0 = 1$, so the $\Gamma$ matrix reduces to the simpler form 
\begin{equation}
\label{eq:TracePreserving_PTM}
    \Gamma = \mqty[1 & \bm{0} \\ \bm{t} & T]\, .
\end{equation}

Furthermore, if the map is also unital (i.e. $\Phi(\mathds{1})=\mathds{1})$), then also $\bm{t} = \bm{0}$.
As an example, the quantum Bit-flip channel described by the map
\begin{equation}
    \mathcal{N}_x(\rho) = (1-p)\rho + p\sigma_x \rho \sigma_x\, , 
\end{equation}
has a corresponding PTM representation as
\begin{equation}
\mathcal{N}_x \longrightarrow \Gamma_x = 
    \begin{bmatrix}
    1 & 0 & 0 & 0 \\
    0 & 1 & 0 & 0 \\
    0 & 0 & 1-2p & 0 \\
    0 & 0 & 0 & 1-2p \\
    \end{bmatrix}\, .
\end{equation}

\subsection{\label{subsec:Deconvolution} Quantum tomographic reconstruction}
Quantum tomography~\cite{DarianoTomography, DARIANO_Tomography, Dariano_Tomography_1, DMacconePaini2003Spin} is a method to estimate the ensemble average of any arbitrary operator by using measurement outcomes of a quorum of observables. The goal of a tomographic reconstruction of an observable is to identify a set of observables $\{Q_\lambda\}$, called \textit{quorum}~\cite{Universal_DAriano}, such that the mean value $\expval{O} = \Tr[O \rho]$ of any observable $O \in \mathcal{L}(\mathcal{H})$, for all states $\rho$, can be reconstructed by using measurements outcomes of the quorum observables. A tomographic reconstruction formula for an operator $O$ is obtained by using a spectral decomposition of the identity in the operator Hilbert space~\cite{Universal_DAriano,ParisQuantumStateEstimation2004,DMacconeParis2001Quorum, Bisio2009OptimalTomography}
\begin{equation}
\label{eq:TomographicRecontructionFormula}
    O = \int_{\Lambda} d\lambda \Tr [C_\lambda^\dagger O] C_\lambda\, ,
\end{equation}
where $\lambda$ is a parameter living in either a continuous or discrete manifold $\Lambda$, and operators $C_\lambda$ depend on the quorum observables. The term $\mathbb{E}[O](Q_\lambda) := \Tr [C^\dagger_\lambda O]\, C_\lambda$ is called \textit{quantum estimator} of the operator $O$, and given a quantum state $\rho$, the expectation value $\expval{O}$ on such state amounts to
\begin{align}
\label{eq:tomography_mean}
\langle O \rangle = \Tr [O\rho]  & =  \int_{\Lambda} d\lambda\, \Tr [OC^\dagger_\lambda]\, \Tr [C_\lambda \rho]\\
& =  \int_{\Lambda} d\lambda\, \Tr [\mathbb{E}[O](Q_\lambda)\, \rho]\\
 & = \int_{\Lambda} d\lambda\, \expval{\mathbb{E}[O](Q_\lambda)}.
\end{align}

For qubit systems, the most common choice (but non unique, e.g.~\cite{DMacconePaini2003Spin}) for the quorum are the Pauli matrices $\{Q_\lambda\}_{\lambda} = \{\sigma_x, \sigma_y, \sigma_z\}$, and the tomographic reconstruction formula results in the standard expansion in the Pauli basis, albeit with a slightly different notation (see Appendix~\ref{app:QubitTomographicFormula} for the explicit derivation):
\begin{equation}
\label{eq:qubit_tomography}
\begin{aligned}
    \expval{O} & = \sum_{\alpha=x,y,z}\frac{1}{3}\, \expval{\mathbb{E}[O](\sigma_\alpha)}\\
    \mathbb{E}[O](\sigma_\alpha) &= \left( \frac{3\Tr[O\sigma_\alpha]}{2}\sigma_{\alpha} + \frac{\Tr[O]}{2}\mathds{1} \right)\, .
\end{aligned}
\end{equation}

Note that the quantum tomographic reconstruction can be straightforwardly applied to  multipartite quantum systems by simply using as a quorum the tensor product of single-system quorums~\cite{Universal_DAriano}.

\section{\label{sec:NoiseDeconvolution} Noise Deconvolution}
The tomographic reconstruction formula can be used whenever one has access to the quantum state $\rho$ and measurements of the quorum observables. In practical scenarios however, estimations are performed in the presence of noise and one generally deals with noisy quantum states $\rho \rightarrow \tilde{\rho} = \mathcal{N}({\rho})$ which then leads to noisy estimates $\expval{O}_{\tilde{\rho}} = \Tr[O\mathcal{N}({\rho})]$. The idea of noise deconvolution is to correct the errors by considering a new quorum of observables taking into account the noise, and then use a noise inverted quantum estimator to recover the ideal estimates, namely the ones that we would obtain in the absence of noise.

Suppose the noise map $\mathcal{N}$ acting on the quantum system can be formally inverted, that is there exist a linear (not CP) map  $\mathcal{N}^{-1}$ such that $(\mathcal{N}^{-1}\circ\mathcal{N})(\rho)=\rho$ for all $\rho$. Then, we say that the noise can be \textit{deconvoluted} in the following sense: instead of measuring the original observable $O$, we can evaluate the expectation value of the noise-inverted operator $\hat{\mathcal{N}}^{-1}(O)$, thus obtaining as a result the desired noise-free ideal result $\expval{O}$, that is
\begin{equation}
\label{eq:noise_inv_informal}
\begin{aligned}
    \expval{\hat{\mathcal{N}}^{-1}(O)}_{\tilde{\rho}} &= \Tr[\hat{\mathcal{N}}^{-1}(O)\,\mathcal{N}(\rho)] \\
    & = \Tr[O\, \mathcal{N}^{-1}(\mathcal{N}(\rho))] \\
    & = \Tr[O\, \rho] = \expval{O}\, ,
\end{aligned}
\end{equation}
where $\hat{\mathcal{N}}^{-1}(\cdot)$ denotes the adjoint of the inverse map $\mathcal{N}^{-1}(\cdot)$, and in the second line we made explicit use of the definition of the adjoint map. 

The condition for \textit{deconvolving} the effect of a noise channel $\mathcal{N}$ at data analysis are~\cite{ParisQuantumStateEstimation2004, Universal_DAriano}:
\begin{itemize}
    \item the inverted noise map exists, that is there is a $\mathcal{N}^{-1}$ such that $(\mathcal{N}^{-1}\circ\mathcal{N})(O) = O \quad \forall O \in \mathcal{L}(\mathcal{H})$. 
    \item the quantum estimator $\mathbb{E}[O](Q_\lambda)$ is in the domain of $\mathcal{N}^{-1}$.
    \item the map $\mathcal{N}^{-1}(\mathbb{E}[O](Q_\lambda))$ is a function of $Q_\lambda$.
\end{itemize}

If these holds, then one can substitute the quantum estimator in Eq.~\eqref{eq:tomography_mean}, with the deconvolved quantum estimator $\hat{\mathcal{N}}^{-1}(\mathbb{E}[O](Q_\lambda))$, yielding
\begin{align}
 \phantom{=} & \int_{\Lambda} d\lambda\, \Tr [\hat{\mathcal{N}}^{-1}(\mathbb{E}[O](Q_\lambda))\, \mathcal{N}(\rho)]\label{eq:inv_noise}\\
 = & \int_{\Lambda} d\lambda\, \Tr [\mathbb{E}[O](Q_\lambda)\, \mathcal{N}^{-1}(\mathcal{N}(\rho))]\nonumber \\
 = & \int_{\Lambda} d\lambda\, \Tr [\mathbb{E}[O](Q_\lambda)\, \rho]\nonumber \\
 = & \Tr[O\rho] = \expval{O}\nonumber\,.
\end{align}

This procedure yields the ideal expectation value of any observable $O$ on the state $\rho$, even if having access only to a noisy version of it and provided that the noise map is known (and invertible). Note that this definition is similar to that recently reported in ref~\cite{Cao_NISQMitigation}, regarding invertible noise channels with non-CPTP inverse. Specializing it for qubits, using Eq.~\eqref{eq:inv_noise} in \eqref{eq:qubit_tomography}, leads to (see Appendix \ref{app:QubitNoiseDeconvolution} for further details)
\begin{equation}
\label{eq:qubit_deconvolution}
    \expval{O} = \frac{1}{2}\Tr[O] + \frac{1}{2}\sum_{\alpha = x,y,z}\Tr[O\sigma_\alpha]\expval{\hat{\mathcal{N}}^{-1}(\sigma_\alpha)}_{\tilde{\rho}}\, .
\end{equation}

Similarly to standard tomographic reconstruction, noise deconvolution can be applied also to multi qubits systems, in which case the mitigated tomographic estimates are obtained considering the tensor product of the deconvolved quantum estimator of each subsystem. Clearly, this holds only for single-qubit noisy channels acting independently on each qubit. In addition, generally non-invertible maps could still be deconvoluted if one restricts the attention only to a subset of states of interest upon which the given map is invertible~\cite{FaithfulStates_Dariano_Presti, Calibration_Dariano_Maccone_Presti}. 

As shown later, the correction of the expectation value of a Pauli matrix is obtained by multiplying the noisy estimate --- the one the experimenter has access to --- by a constant depending on the noise, i.e. $\expval{\sigma_\alpha}_{mitig} = c\expval{\sigma_\alpha}_{noisy}$. This clearly increases the variance of the estimation, since $\text{Var}[\expval{\sigma_\alpha}_{mitig}] = c^2 \text{Var}[\expval{\sigma_\alpha}_{noisy}] \sim c^2 / M$, where $M$ is the number of measurements performed on the system, and thus the experimenter need to increase the outcome statistics proportionally to $c^2$ to reach a desired target precision.

We now proceed discussing how the deconvolution behaves in the presence of multiple noise channels. Consider two noise processes $\mathcal{N}_0$ and $\mathcal{N}_1$ separated by a unitary gate $\mathcal{U}(\cdot) = U \cdot U^\dagger$, as shown in Fig.~\ref{fig:noise_deconvolution_summary}(d). The action of the circuit is
\begin{align}
    \big(\mathcal{N}_1 \circ \mathcal{U} \circ \mathcal{N}_0 \big)(\rho)
    & = \mathcal{N}_1 \bigg(U \mathcal{N}_0 \big(\rho\big)U^\dagger\bigg)\nonumber \\
    & = \mathcal{N}_1 (\tilde{\rho}_{U})\nonumber
\end{align}
with $\tilde{\rho}_{U} = U\mathcal{N}_0 \big(\rho\big)U^\dagger$. Using~\eqref{eq:inv_noise}, it is possible to deconvolve the outermost noise $\mathcal{N}_1$ with 
\begin{equation}
    \Tr[\hat{\mathcal{N}}^{-1}_1(O)\mathcal{N}_1(\tilde{\rho}_U)]\,,
\end{equation}
but not $\mathcal{N}_0$, since the unitary $U$ is in the way. Actually, one could decide to deconvolve the unitary as well, using the trivial inverse $\mathcal{U}_1^{-1}(\cdot) = U_1^\dagger \cdot U_1$, and thus making it possible to deconvolve also the first noise channel $\mathcal{N}_0$, as 
\begin{align}
    & \Tr[\hat{\mathcal{N}}^{-1}_1\big((U\hat{\mathcal{N}}^{-1}_0(O)U^\dagger\big)\mathcal{N}_1(\tilde{\rho}_U)]\, \nonumber\\
    &= \Tr[\big(U\hat{\mathcal{N}}^{-1}_0(O)U^\dagger\big) \,U\mathcal{N}_0\big(\rho\big)U^\dagger)] \nonumber\\
    &= \Tr[\hat{\mathcal{N}}^{-1}_0(O)\mathcal{N}_0(\rho)]\nonumber\\
    &= \Tr[O\rho]\, \nonumber.
\end{align}
However, this procedure cannot be employed to invert the noise that happens before a generic unitary $U$, since it essentially offloads the computation from the quantum computer to the classical one, by simulating the inverse evolution of the quantum system.

A more interesting case of is obtained when the error map happens to commute with all the remaining operation in the computation, as is the case for the depolarizing noise, described by the map
\begin{equation}
    \mathcal{N}_{\text{dep}}(\rho) = \frac{p\mathds{1}}{2} + (1-p)\rho\, ,
\end{equation}
for which it is easy to see that $\big(\mathcal{N}_{\text{dep}}\circ \mathcal{U}\big)(\rho) = \big(\mathcal{U}\circ \mathcal{N}_{\text{dep}}\big)(\rho)\, \forall\,\, \mathcal{U}(\cdot)=U\cdot U^\dagger$. Suppose one is performing a quantum computation given by a sequence of operations $U_i$, each one followed by depolarizing noise
\begin{align}
    \rho & = \bigg(\prod_{i=1}^{d} \mathcal{N}^{(i)}_{\text{dep}}\circ \mathcal{U}_i\,\bigg) (\rho_0) \\
    & = \bigg(\prod_{i=1}^{d}\mathcal{N}^{(i)}_{\text{dep}} \circ \prod_{i=1}^{d}\mathcal{U}_i\bigg)(\rho_0) \\
    & = \mathcal{N}_{\text{dep}}^{\text{tot}}(\rho_U)\, ,
\end{align}
with $\mathcal{N}_{\text{dep}}^{\text{tot}} = \prod \mathcal{N}_{\text{dep}}^{(i)}$ the composition of all the depolarizing channels, and $\rho_U = \prod \mathcal{U}_i(\rho_0)$ the state obtained by the ideal noise-free computation. Most importantly, one can check that the composition of multiple depolarizing channels is still a depolarizing channel with probability parameter $1-p_{\text{tot}} = \prod(1-p_i)$, where $p_i$ is the probability associated with each depolarizing noise. In such case it is possible to deconvolve all noise at once, using the deconvolution formula for the depolarizing noise with the total noise parameter $p_\text{tot}$ (see Eq.~\eqref{eq:depol_deconv}). Similarly, this also holds for computations involving multi qubits subject to \textit{global} depolarizing errors. The authors in ref.~\cite{GlobalDepolarizingInversion} leverage this property to perform a simple yet effective error mitigation technique for quantum computers, based on the assumption that noise in quantum circuits is well described by global depolarizing error channels. While exact depolarizing errors (either local or global) are hardly found in realistic quantum circuits where errors are both due to \textit{coherent} (i.e. unitary) and \textit{incoherent} noise (i.e. interaction), Pauli twirling and randomized compiling techniques~\cite{RandomizedCompiling, RandomizedCompilingExperiments, RB, PauliRandomizationRC_Ware} can be used to approximately tailor noise to stochastic Pauli channels, preferably depolarizing noise, and then use the procedure above to mitigate it~\cite{Ville2021TailoringDepolarizing}.

\section{\label{sec:MapInversion} Inversion of common noise maps}

\begin{table*}
\begin{ruledtabular}
\begin{tabular}{lcc}
    & Noise Channel $\mathcal{N}(\rho)$ & Inverse Map $\mathcal{N}^{-1}(O)$ \\ \hline\\[0pt]
    
    Bit-Flip   &  $(1-p)\rho + p\sx\rho \sx$   &   $\displaystyle{\frac{1-p}{1-2p}O - \frac{p}{1-2p}\sx O \sx}$ \\[10pt]
    
    Phase-Flip (or \textit{dephasing})   &  $(1-p)\rho + p\sz\rho \sz$   &  $\displaystyle{\frac{1-p}{1-2p}O - \frac{p}{1-2p}\sz O \sz}$ \\[10pt]
    
    Bit-Phase-Flip  &  $(1-p)\rho + p\sy\rho \sy$   &   $\displaystyle{\frac{1-p}{1-2p}O - \frac{p}{1-2p}\sy O \sy}$ \\[10pt]
    
    Depolarizing    &  $(1-p)\rho + p\displaystyle{\frac{\mathds{1}}{2}}$  &     $\displaystyle{\frac{1}{1-p}\left(O-\frac{p}{2}\Tr[O]\mathds{1} \right)}$ \\[10pt]
    
    General Pauli Channel   &  $p_0 \rho + p_x \sx\rho \sx + p_y \sy\rho \sy + p_z \sz\rho \sz$ & $\beta_0 O + \beta_1 \sx O \sx + \beta_2 \sy O \sy + \beta_3 \sz O \sz$ \\[0pt] 
     & & (see Eq.~\eqref{eq:inv_paulichannel} for the coefficients) \\[10pt]
    
    Amplitude Damping & $V_0\rho V_0 + V_1 \rho V_1^\dagger$  & $K_0 O K_0 - K_1 O K_1^\dagger$ \\[0pt]
     & $\begin{array} {lcl} & V_0 = \dyad{0} + \sqrt{1-\gamma}\dyad{1} \\[5pt] & V_1 = \sqrt{\gamma}\ketbra{0}{1}\end{array}$ & $\begin{array} {lcl} & K_0 = \dyad{0} + \sqrt{\frac{1}{1-\gamma}}\dyad{1} \\[5pt] & K_1 = \sqrt{\frac{\gamma}{1-\gamma}}\ketbra{0}{1}\end{array}$\\[20pt]
    
    2-Kraus Channel   & $A_0\rho A_0 + A_1 \rho A_1^\dagger$  &  $B_1 O B_1^\dagger - B_2 O B_2^\dagger$ \\[0pt]
     & $\begin{array} {lcl} & A_0 = \cos\alpha\dyad{0} + \cos\beta\dyad{1}\\[5pt] & A_1 = \sin\beta\ketbra{0}{1}+\sin\alpha\ketbra{1}{0}\end{array}$ & $\begin{array} {lcl} & B_0 = \frac{\sqrt{2}\cos\beta}{\sqrt{\cos2\alpha+\cos2\beta}}\dyad{0} + \frac{\sqrt{2}\cos\alpha}{\sqrt{\cos2\alpha+\cos2\beta}}\dyad{1} \\[5pt] & B_1 = \frac{\sqrt{2}\sin\beta}{\sqrt{\cos2\alpha+\cos2\beta}}\ketbra{0}{1}+\frac{\sqrt{2}\sin\alpha}{\sqrt{\cos2\alpha+\cos2\beta}}\ketbra{1}{0}\end{array}$ \\[20pt]
\end{tabular}
\end{ruledtabular}
\caption{\label{tab:tablesummary} Table summarizing the results of the present analysis, consisting of some of the most common single-qubit noisy channels $\mathcal{N}$, along with their inverse noise maps $\mathcal{N}^{-1}$, defined as the map such that $(\mathcal{N}^{-1}\circ\mathcal{N})(\rho) = \rho\,\, \forall\, \rho$. Clearly, all noise channels are CPTP maps, while the inverse channels are not, yet they admit an operator-sum representation. All the noise maps except for amplitude damping and 2-Kraus channel have trivial adjoint channels, so one must pay attention in using the adjoint channel inside the deconvolution formula~\eqref{eq:qubit_deconvolution}.}
\end{table*}

We now proceed by explicitly evaluating the inverse maps of some of the most common noisy channels, leveraging the Pauli Transfer Matrix formalism introduced in Sec~\ref{sec:Preliminaries}. The general method for finding the inverse map goes as follows: we first evaluate the matrix representation~\eqref{eq:matrix_channel} of the channel, we then invert this matrix, and from this recover the operator sum representation of the inverse channel whenever this exists. We start from simpler cases to build some intuition on the construction of the inverse maps, and then proceed towards more complicated cases. In Table~\ref{tab:tablesummary} we summarize the results obtained in this section, comprising all noise channels considered in this analysis together with their inverse maps. 

\subsubsection{Bit-flip, Phase-flip and Bit-Phase-flip}
The Bit-flip, Phase-flip and Bit-Phase-flip channels are described by the Kraus operators, $A_0 = \sqrt{p}\mathds{1}$ and $A_{1,\alpha}$ = $\sqrt{1-p}\,\sigma_\alpha$, with $\sigma_\alpha \in \{\sigma_x,\sigma_z,\sigma_y\}$ respectively. For simplicity, in the following we focus only on the Bit-flip channel (generated by $\sigma_x$), but the results hold equivalently also for the other two channels. The Bit-flip channel acts as:
\begin{equation}
\label{eq:bit-flip}
\mathcal{N}_x(\rho) = (1-p)\rho + p\sigma_x\rho \sigma_x
\end{equation}
and its PTM is given by 
\begin{equation}
\Gamma_x = 
    \begin{bmatrix}
    1 & 0 & 0 & 0 \\
    0 & 1 & 0 & 0 \\
    0 & 0 & 1-2p & 0 \\
    0 & 0 & 0 & 1-2p \\
    \end{bmatrix}\, .
\end{equation}

In order to find an operator sum expression for the inverse map $\mathcal{N}_x^{-1}$, we consider the inverse matrix $\Gamma_x^{-1}$
\begin{equation}
\Gamma_x^{-1} = 
    \begin{bmatrix}
    1 & 0 & 0 & 0 \\
    0 & 1 & 0 & 0 \\
    0 & 0 & \frac{1}{(1-2p)} & 0 \\
    0 & 0 & 0 & \frac{1}{(1-2p)} \\
    \end{bmatrix}\, .
\end{equation}
It's clear that $\Gamma_x$ can be inverted provided that $p\neq 1/2$, since in that case $\text{det}\,\Gamma_x=0$. This is not a problem for realistic case scenarios, where the probability of errors are small, so that one can safely assume $0<p<1/2$. We now proceed using a similar procedure found in~\cite{BourdonUnitalQuantumOperations2004}.

Note that $\Gamma_x^{-1}$ is diagonal in the Pauli basis, thus has eigenvectors $\{|\mathds{1}\rrangle, |\sigma_x\rrangle\,|\sigma_y\rrangle\,|\sigma_z\rrangle\}$ with eigenvalues $\bm{\lambda} = \{1, 1, (1-2p)^{-1}, (1-2p)^{-1}\}$ respectively. Now, consider the generic map 
\begin{equation}
\label{eq:op-sum}
    \mathcal{E}(O) = \sum_{j=0}^{3}\beta_j \sigma_j O \sigma_j\, .
\end{equation}
Also this map has eigenvectors $\{\mathds{1}, \bm{\sigma}\}$, but with eigenvalues $\bm{\beta}=\{\beta_0, \beta_1, \beta_2, \beta_3\}$. Since two maps are equal if they have the same action on a basis, if we can find a way to match the two sets of eigenvalues $\bm{\lambda}$ and $\bm{\beta}$, we would then recover the operator-sum representation for $\Gamma_x^{-1}$. 

By evaluating the PTM $\Gamma_{\mathcal{E}}$ of $\mathcal{E}$, we can relate the coefficients in the operator-sum representation~\eqref{eq:op-sum}, with those appearing in the expression for $\Gamma_x^{-1}$ (see Appendix~\ref{app:inverse_maps} for a derivation). In particular, we want these to hold:
\begin{align}
(\Gamma_x^{-1})_{11} = \beta_0 + \beta_1 - \beta_2 - \beta_3\\
(\Gamma_x^{-1})_{22} = \beta_0 - \beta_1 + \beta_2 - \beta_3\\
(\Gamma_x^{-1})_{33} = \beta_0 - \beta_1 - \beta_2 + \beta_3
\end{align}
plus the trace-preserving condition $1 = \beta_0 + \beta_1 + \beta_2 + \beta_3$, that the inverse map must satisfy because the direct map is trace-preserving. This condition is inherently satisfied by $\Gamma_x^{-1}$, since its first row has the form $(1, 0, 0, 0)$. This system has solutions $\beta_0 = (1-p)/(1-2p)$, $\beta_1=-p/(1-2p)$, and $\beta_2=\beta_3=0$, and substituting them back into Eq.~\eqref{eq:op-sum}, we obtain the operator-sum representation of the inverse Bit-flip map
\begin{equation}
\label{eq:bit-flip-inv}
    \mathcal{N}_x^{-1}(O) = \frac{1-p}{1-2p}O - \frac{p}{1-2p}\sigma_x O \sigma_x\, .
\end{equation}
By virtue of Corollary II, and noticing that the coefficients appearing in the expression above have always opposite signs, we are sure that this map is not CP, as expected, yet it possesses an operator sum representation. Note how similar the direct and inverse map are, a feature which we will encounter in all the cases discussed here. 

The same procedure can be applied to Phase-flip (or \textit{dephasing}, generated by $\sigma_z$), and Bit-Phase-flip (generated by $\sigma_y$) channels, yielding inverse maps
\begin{align}
    \mathcal{N}_z^{-1}(O) & = \frac{1-p}{1-2p}O - \frac{p}{1-2p}\sigma_z O \sigma_z\\
    \mathcal{N}_y^{-1}(O) & = \frac{1-p}{1-2p}O - \frac{p}{1-2p}\sigma_y O \sigma_y\, .
\end{align}

We can plug these inversion maps in the deconvolution formula~\eqref{eq:qubit_deconvolution} to obtain noise-free expectation values. In particular, assume we are measuring a Pauli matrix $O=\sigma_\alpha$, and that the system is subject to one of the noise processes $\rho \rightarrow \rho_\beta = \mathcal{N}_\beta(\rho)$ with $\beta = \{x,y,z\}$. Then the ideal expectation values $\expval{\sigma_\alpha}_{\rho} = \Tr[\sigma_\alpha\rho]$ can be expressed in compact form as (see Appendix~\ref{app:inverse_maps} for the explicit derivation)
\begin{equation}
\label{eq:pauli_deconv}
    \expval{\sigma_\alpha}  = \delta_{\alpha \beta}\, \expval{\sigma_\alpha}_{\rho_\beta} + (1-\delta_{\alpha \beta})\,  \frac{1}{1-2p}\, \expval{\sigma_\alpha}_{\rho_\beta}\, .
\end{equation}
It is then clear that if the noise happens along the measurement direction ($\alpha=\beta$), then the noise does not affect the measurement statistics, as the ideal and noisy value coincide. While for orthogonal directions ($\alpha\neq\beta$), these are equally contracted by a factor $1-2p$, thus recovering the usual pictorial representation of the contracting Bloch sphere on the plane orthogonal to the noise~\cite{NielsenChuang}.

\subsubsection{Depolarizing noise}
The depolarizing noise, introduced above,
\begin{equation}
    \mathcal{N}_{\text{dep}}(\rho) = (1-p)\rho + \frac{p\mathds{1}}{2}\, ,
\end{equation}
leaves the state untouched with probability $1-p$, and sends it to the completely mixed state $\mathds{1}/2$ with probability $p$. The channel can be expressed in Kraus form in multiples ways, one of them being~\cite{NielsenChuang}
\begin{equation}
    \mathcal{N}_{\text{depol}}(\rho) = \bigg(1-\frac{3p}{4}\bigg)\rho + \frac{p}{4}\bigg( \sigma_x \rho \sigma_x + \sigma_y \rho \sigma_y + \sigma_z \rho \sigma_z\bigg)\, ,
\end{equation}
with Kraus operators $\{A_0 = \sqrt{1-3p/4}\,\mathds{1},\, A_1 = \sqrt{p}\,\sigma_x/2\,, A_2 = \sqrt{p}\,\sigma_y/2\,, A_3 = \sqrt{p}\,\sigma_z/2 \}$.
Following the same procedure used for Bit-flip channel, one obtains the inverse linear map (see Appendix~\ref{app:inverse_maps})
\begin{equation}
\label{eq:depol_inv}
\mathcal{N}_{\text{depol}}^{-1}(O) = \frac{1}{1-p}\left( O-\frac{p}{2}\Tr[O]\mathds{1} \right)\, .
\end{equation}

While this is already a known result in the literature~\cite{DarianoDepolarizingInversion, Bisio2009OptimalTomography, Universal_DAriano, huangPredictingManyProperties2020a, Temme_PEC}, it is presented without an explicit constructive derivation, as given here. 

Using this formula in the deconvolution tomographic reconstruction~\eqref{eq:qubit_deconvolution}, we find
\begin{equation}
\label{eq:depol_deconv}
    \expval{O} = \frac{1}{2}\Tr[O] + \sum_{\alpha}\frac{\Tr[O\sigma_\alpha]}{1-p}\expval{\sigma_\alpha}_{\mathcal{N}_{\text{dep}}(\rho)}\, ,
\end{equation}
where it is clear that to counterbalance the effect of the depolarizing channel, whose effect on the Bloch sphere is to contract it uniformly, one needs perform an expansion of the same amount, obtained dividing by $1-p$.

\subsubsection{General Pauli channel}
A more general and interesting case is the general Pauli channel, where noise intensities are different along the three Pauli axes
\begin{equation}
\label{eq:general_pauli_channel}
\mathcal{N}_{\bm{p}}(\rho) = p_0 O + p_x \xox{\rho} + p_y \yoy{\rho} + p_z \zoz{\rho}\, .
\end{equation}
The channel is parametrized by the probabilities $\boldsymbol{p}=(p_0, p_x, p_y, p_z)$, with the trace-preserving condition implying $p_0 = 1-p_x-p_y-p_z$. Upon choosing appropriate values for $\boldsymbol{p}$, this channel reduces to all noise maps treated before. Though of considerable more general structure, the inverse map of this channel is derived using the same machinery developed above, and eventually one obtains

\begin{widetext}
\begin{equation}
\label{eq:inv_paulichannel}
\begin{alignedat}{1}
& \mathcal{N}^{-1}_{\bm{p}}(O) = \beta_0 O + \beta_1 \xox{O} + \beta_2 \yoy{O} + \beta_3 \zoz{O}\quad \text{with}\\
& \beta_0 =\frac14\bigg(1+\frac{1}{1-2(p_x+p_y)}+\frac{1}{1-2(p_x+p_z)}+\frac{1}{1-2(p_y+p_z)}\bigg)\\
& \beta_1 = \frac14\bigg(1-\frac{1}{1-2(p_x+p_y)}-\frac{1}{1-2(p_x+p_z)}+\frac{1}{1-2(p_y+p_z)}\bigg)\\ 
& \beta_2 = \frac14\bigg(1-\frac{1}{1-2(p_x+p_y)}+\frac{1}{1-2(p_x+p_z)}-\frac{1}{1-2(p_y+p_z)}\bigg) \\
& \beta_3 = \frac14\bigg(1+\frac{1}{1+2(p_x+p_y)}-\frac{1}{1-2(p_x+p_z)}-\frac{1}{1-2(p_y+p_z)}\bigg)\, .
\end{alignedat}
\end{equation}
\end{widetext}

One can check that varying $\bm{p}$ it is possible to recover the inverse maps of all the cases treated before. For example, for $\bm{p} = (1-p, p, 0, 0)$ corresponding to the bit-flip channel, one gets $\beta_0 = (1-p)/(1-2p)$ and $\beta_1=-p/(1-2p)$, as in Eq.~\eqref{eq:bit-flip-inv}.

The noise deconvolution applied to measurements of Pauli matrices $O \in \{\sigma_x, \sigma_y, \sigma_z\}$, leads to the following relations
\begin{equation}
\label{eq:deconvolution_gpc}
\begin{aligned}
    \expval{\sigma_x} & = \frac{1}{1-2(p_y+p_z)}\expval{\sigma_x}_{\mathcal{N}_{\bm{p}}(\rho)}\\
    \expval{\sigma_y} & = \frac{1}{1-2(p_x+p_z)}\expval{\sigma_y}_{\mathcal{N}_{\bm{p}}(\rho)}\\
    \expval{\sigma_z} & = \frac{1}{1-2(p_x+p_y)}\expval{\sigma_z}_{\mathcal{N}_{\bm{p}}(\rho)}\,,
\end{aligned}
\end{equation}

which can be used together with~\eqref{eq:qubit_deconvolution} to reconstruct the expectation value of a general observable $O$. Again, we see that the noise disturbs the estimation along orthogonal directions, as in all previous cases. Note that the explicit inversion of the general Pauli channel was also recently reported in ref.~\cite{suzuki2021quantum}.

\subsubsection{Amplitude Damping}
The amplitude damping (AD) channel describes the energy loss of a quantum system, for example obtained through relaxation from the excited to the ground state. Its Kraus representation is
\begin{equation}
\label{eq:AmplitudeDamping}
\begin{aligned}
    & \mathcal{N}_{\text{AD}}(\rho) = V_0\rho V_0^\dagger + V_1\rho V_1^\dagger\, ,\\
    & K_0 = \begin{bmatrix}
    1 & 0 \\ 
    0 & \sqrt{1-\gamma}
    \end{bmatrix}\quad
    K_1 = \begin{bmatrix}
    0 & \sqrt{\gamma}\\
    0 & 0
    \end{bmatrix}\, .
\end{aligned}
\end{equation}

While still being trace preserving (TP), amplitude damping channel is not unital, since $\mathcal{N}_{\text{AD}}(\mathds{1}) = \mathds{1}+\gamma Z$. This in turn implies that the Pauli Transfer Matrix $\Gamma_{\text{AD}}$ is not diagonal, but has an addition nonzero element in the last row of first column. This changes the derivation of the inverse map with respect to the previous cases, but it can still be carried out without major changes (see Appendix~\ref{app:AmplitudeDamping}). The inverse linear map in operator sum representation is then found to be
\begin{equation}
\label{eq:AD_inv}
\begin{aligned}
& \mathcal{N}_{\text{AD}}^{-1}(\rho) = K_0 O K_0^\dagger - K_1 O K_1^\dagger\, ,\\
& K_0 = 
\begin{bmatrix}
1 & 0 \\
0 & \frac{1}{\sqrt{1-\gamma}}
\end{bmatrix}
\,, 
K_1 = 
\begin{bmatrix}
0 & \sqrt{\frac{\gamma}{1-\gamma}} \\ 
0 & 0
\end{bmatrix}\, .
\end{aligned}
\end{equation}

Up until now, all noisy channels (and their inverse maps) had trivial \textit{adjoint} map, since all Kraus operators were Hermitian. However this is not the case for amplitude damping, since both $V_1\neq V_1^\dagger$ and $K_1\neq K_1^\dagger$. Thus, one must be careful in applying the adjoint inverse $\hat{\mathcal{N}}^{-1}$ in Eq.~\eqref{eq:qubit_deconvolution}, and not just $\mathcal{N}^{-1}$ of \eqref{eq:AD_inv} (see Appendix~\ref{app:AmplitudeDamping} for an extended discussion).
Deconvolution of amplitude damping for measurements of the Pauli matrices leads to 
\begin{equation}
\label{eq:ad_sigmas}
\begin{aligned}
\expval{\sigma_x} & = \frac{1}{\sqrt{1-\gamma}}\langle \sigma_x \rangle_{\mathcal{N}_{\text{AD}}(\rho)} \\
\expval{\sigma_y} & = \frac{1}{\sqrt{1-\gamma}}\langle \sigma_y \rangle_{\mathcal{N}_{\text{AD}}(\rho)} \\
\expval{\sigma_z} & = \frac{1}{1-\gamma}\qty(\langle \sigma_z \rangle_{\mathcal{N}_{\text{AD}}(\rho)} -\gamma)\, .
\end{aligned}
\end{equation}

\subsubsection{Two-Kraus channels}
We now move our attention to the set of channels generated by two parametrized Kraus operators
\begin{equation}
\mathcal{N}_{\text{two}}(\rho) = \sum_{i=1, 2}A_i\rho A_i^\dagger\, ,  
\end{equation}
with $A_1 = \cos\alpha\dyad{0}+\cos\beta\dyad{1}$, and $A_2 = \sin\beta\ketbra{0}{1}+\sin\alpha\ketbra{1}{0}$. This channel reduces to Bit-flip for $\alpha=\beta$, and to amplitude damping for $\alpha=0$.  

Following a procedure similar to the amplitude damping case, the inverse map of the two-Kraus channels is found to be
\begin{equation}
\begin{aligned}
& \mathcal{N}_{\text{two}}(O)^{-1} = B_1 O B_1^\dagger - B_2 O B_2^\dagger \quad \text{with}\\
&
B_1 = \begin{bmatrix}
\frac{\sqrt{2}\cos\beta}{\sqrt{\cos2\alpha+\cos2\beta}} & 0 \\ 
0 & \frac{\sqrt{2}\cos\alpha}{\sqrt{\cos2\alpha+\cos2\beta}}
\end{bmatrix} \\
& 
B_2 = \begin{bmatrix}
0 & \frac{\sqrt{2}\sin\beta}{\sqrt{\cos2\alpha+\cos2\beta}} \\
\frac{\sqrt{2}\sin\alpha}{\sqrt{\cos2\alpha+\cos2\beta}} & 0
\end{bmatrix}
\end{aligned}
\end{equation}

Similarly to amplitude damping, one of the generators ($B_2$) is not Hermitian, thus we must employ the adjoint inverse map when evaluating the deconvolved expectation values. By straightforward calculations the following hold:
\begin{equation}
\begin{aligned}
\expval{\sigma_x} & = \frac{1}{\cos(\alpha-\beta)}\expval{\sigma_x}_{\mathcal{N}_{\text{two}}(\rho)} \\
\expval{\sigma_y} & = \frac{1}{\cos(\alpha+\beta)}\expval{\sigma_y}_{\mathcal{N}_{\text{two}}(\rho)} \\
\expval{\sigma_z} & = h_{\alpha\beta}\big(\cos^2\beta+\sin^2\alpha-1+\expval{\sigma_z}_{\mathcal{N}_{\text{two}}(\rho)}\big)\, ,
\end{aligned}
\end{equation}
with $h_{\alpha\beta} = \frac{2}{\cos(2\alpha)+\cos(2\beta)}$.

Note that upon varying the parameters $\alpha$ and $\beta$, the formulas above correctly reduce to the other limiting channels. For example, setting $\alpha=0$ leads to amplitude damping channel~\eqref{eq:ad_sigmas}, with $\cos(\beta):=\sqrt{1-\gamma}$.

\section{\label{sec:Experiments} Experimental deconvolution}

In this section we provide some concrete applications of the noise deconvolution procedures for qubit tomography outlined above. In particular, we show both numerically and by experimentation on quantum hardware by Rigetti how to address a \textit{dechoerence} noise model, and we also provide numerical evidence for the deconvolution of the general Pauli channel~\eqref{eq:general_pauli_channel}. All simulations are performed using PyQuil and the real quantum device usded is ``Aspen-9", accessed via Rigetti's Quantum Cloud Services (QCS)~\cite{smith2016practical, Karalekas_2020}. 

\subsection{Decoherence noise model}
\begin{figure*}[htbp]
    \centering
    \includegraphics[width = \textwidth]{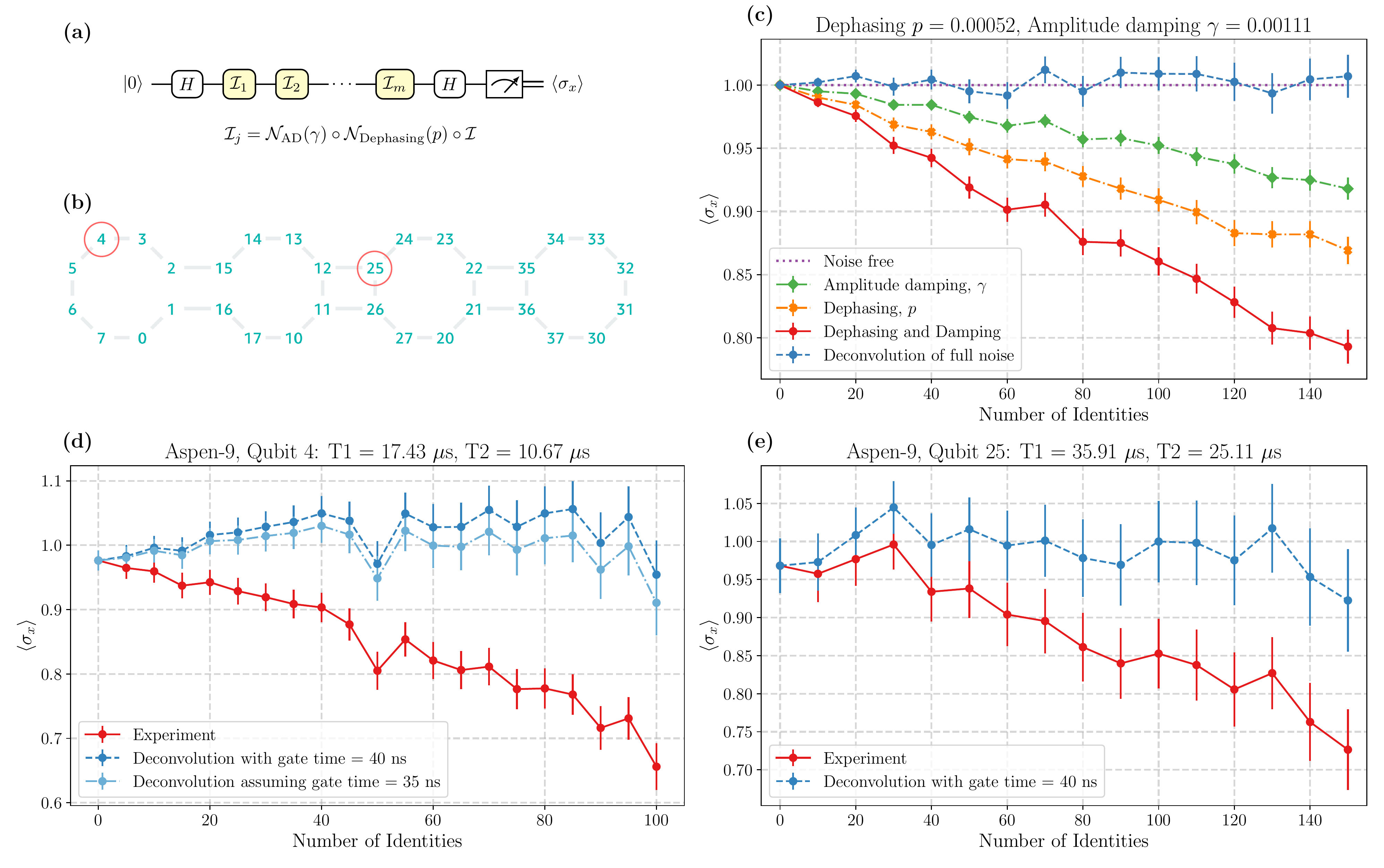}
    \caption{Deconvolution of decoherence noise both on a simulator and the real quantum device Aspen-9 by Rigetti. \textbf{(a)} Scheme of the quantum circuit used in the simulations and runs on the actual quantum device. A qubit is prepared in the superposition state and then it is left to decohere for a certain amount of time, dependent on the number $m$ of identities in the circuit. Eventually the qubit is measured in the $x$ basis to estimate the expectation value $\sigma_x$. \textbf{(b)} Scheme of Aspen-9, the real quantum device by Rigetti used to run the quantum circuit.  \textbf{(c)} Simulation of the decoherence noise for dephasing ($p$) and damping ($\gamma$) intensities equal to those characterizing qubit $25$ of Aspen-9, with gate duration of $40$~ns. For comparison, the effect of the action of these channels alone are also showed. Using the deconvolution formulas for decoherence noise~\eqref{eq:decoherence_sigmas_multiple}, it is possible to mitigate the decay caused by the noise, and recover the ideal result. Each circuit is sampled with $n_{\text{shots}} = 2048$ shots. \textbf{(d)} Results obtained from running the circuit on qubit $4$ of Aspen-9, characterized by relaxation times $T_1 = 17.43\cdot 10^{-6}$~s and $T_2 = 10.67\cdot 10^{-6}$~s, with $n_{\text{shots}} = 2048$. See main text for comments on the results. \textbf{(e)} Results obtained from running the circuit on qubit $25$ of Aspen-9, characterized by relaxation times $T_1 = 35.91\cdot 10^{-6}$~s and $T_2 = 25.11\cdot 10^{-6}$~s, with $n_{\text{shots}} = 1024$. See main text for comments on the results.}
    \label{fig:decoherence_noise}
\end{figure*}

The concurrent action of a dephasing channel followed by amplitude damping is referred to as \textit{decoherence} noise, which is an effective way to describe the noisy evolution a qubit undergoes due to uncontrolled interaction with its external environment. Using the definitions~\eqref{eq:bit-flip} and \eqref{eq:AmplitudeDamping}, one obtains
\begin{equation}
\begin{aligned}
\mathcal{N}_{\text{dec}}(\rho) & = \big(\mathcal{N}_{\text{AD}}(\gamma) \circ \mathcal{N}_z (p)\big)\qty(\mqty[a & b \\ c & 1-a]) \\
& = \mqty[ 1-(1-a)(1-\gamma) & (1-2p)\sqrt{1-\gamma}\,b \\ (1-2p)\sqrt{1-\gamma}\,c & (1-\gamma)\,(1-a)] \\
& = \mqty[1-(1-a)e^{-t/T_1} & e^{-t/T_2}\,b \\ e^{-t/T_2}\,c & e^{-t/T_1}\,(1-a)]\,,
\end{aligned}
\end{equation}

where we have introduced the relaxation times $T_1$ and $T_2$ characterizing the ``quality" of the physical qubits. These are related to the noise parameters $\gamma$ and $p$ through the following relations
\begin{equation}
\label{eq:noise_and-Ts}
\begin{aligned}
    \gamma &= 1-e^{-t/T_1}\\
    p &= \frac{1}{2}\big(1-e^{-(t/T_2-t/2T_1)}\big)\, ,
\end{aligned}
\end{equation}
where $t$ is a time parameter indicating the duration of the noise process. 

Since the correction terms in the deconvolution formulas for dephasing~\eqref{eq:ad_sigmas} and amplitude damping\eqref{eq:pauli_deconv} are multiplicative, for a decoherence channel these combine as
\begin{equation}
\label{eq:decoherence_sigmas}
\begin{aligned}
\expval{\sigma_x} & = \frac{1}{(1-2p)}\frac{1}{\sqrt{1-\gamma}}\langle \sigma_x \rangle_{\mathcal{N}_{\text{dec}}(\rho)} \\
\expval{\sigma_y} & = \frac{1}{(1-2p)}\frac{1}{\sqrt{1-\gamma}}\langle \sigma_y \rangle_{\mathcal{N}_{\text{dec}}(\rho)} \\
\expval{\sigma_z} & = \frac{1}{1-\gamma}\qty(\langle \sigma_z \rangle_{\mathcal{N}_{\text{dec}}(\rho)} -\gamma)\, .
\end{aligned}
\end{equation}

Additionally, if the quantum system under investigation is subject to repeated applications of a decoherence noise channel, i.e. $\mathcal{N}_{\text{dec}}^{\circ m}(\rho) =\mathcal{N}^{(1})_{\text{dec}}\circ \mathcal{N}^{(2)}_{\text{dec}} \cdots \circ \mathcal{N}^{(m)}_{\text{dec}}(\rho)$, then the ideal expectation values are obtained through the following equations
\begin{equation}
\label{eq:decoherence_sigmas_multiple}
\begin{aligned}
\expval{\sigma_x} & = \frac{1}{\qty((1-2p)\sqrt{1-\gamma})^m}\langle \sigma_x \rangle_{\mathcal{N}_{\text{dec}}(\rho)} \\
\expval{\sigma_y} & = \frac{1}{\qty((1-2p)\sqrt{1-\gamma})^m}\langle \sigma_y \rangle_{\mathcal{N}_{\text{dec}}(\rho)} \\
\expval{\sigma_z} & = \frac{1}{(1-\gamma)^m}\qty(\langle \sigma_z \rangle_{\mathcal{N}_{\text{dec}}(\rho)} -1+(1-\gamma)^m)\, .
\end{aligned}
\end{equation}

In Figure~\ref{fig:decoherence_noise} we show the application of these formulas to deconvolve the decoherence noise occurring on a qubit. The specific quantum circuit used for the experiments is showed in Figure~\ref{fig:decoherence_noise}(a): first the system is prepared in the superposition state $\ket{+}=H\ket{0}=(\ket{0}+\ket{1})/\sqrt{2}$, then we let qubit decohere for a certain amount of time dictated by the number $m$ of (noisy) identities each of which takes a time $t$, and at last we measure the expectation value of the operator $\sigma_x$. Clearly, in a noise-free scenario, the result would always be $\expval{\sigma_x}=1$, independent of the depth $m$. Figure~\ref{fig:decoherence_noise}(c) shows a simulation of these circuits with stochastic measurement outcomes for different values of $m$, and for a given choice of noise parameters $p$ and $\gamma$. For comparison, the individual effect of dephasing and amplitude damping channels alone are also showed. Thanks to Eq.~\eqref{eq:decoherence_sigmas_multiple} we can invert the effect of the decoherence noise, and so retrieve the ideal noise-free results.

We also tested this procedure on quantum hardware provided by Rigetti, in particular on the device ``Aspen-9", whose topology is reported in Fig.~\ref{fig:decoherence_noise}(b). The device comes with the calibration data reporting the $T_1$ and $T_2$ parameters for any qubit, as well as the time duration of a single gate. Identities in the circuits are used introduce time delays, and thus let the qubit decohere for longer intervals of time, depending on the depth $m$. Differently from the previous simulations where only the identities are supposed to introduce (decoherence) noise, in the real case scenario noise happens along the whole computation, including state preparation, application of all gates in the circuit (both Hadamards and Identities), and finally measurement errors. Of these, the most detrimental are undoubtedly readout errors, and we addressed them by using the standard mitigation technique of calibrating the device and inverting the assignment probability matrix to recover readout mitigated results. Calibration data reports that the time it takes to execute a single qubit identity gate is $t=40$~ns, and together with $T_1$ and $T_2$, these are used to calculate the parameters $p$ and $\gamma$ of the decoherence noise, using relations~\eqref{eq:noise_and-Ts}. These are in turn used inside the deconvolution formulas to recover the noise-free results. Figures~\ref{fig:decoherence_noise}(d) and \ref{fig:decoherence_noise}(e) show the results of the execution of circuit  Fig.~\ref{fig:decoherence_noise}(a) on qubits $4$ and $25$, respectively. 

The noise mitigation procedure on qubit $4$ shown in panel~\ref{fig:decoherence_noise}(d), yields slightly unphysical results, in the sense that the mitigated value exceeds one, which is of course not possible. A naive solution to this problem could be to impose that the mitigation results are in the physical range $[-1,+1]$, so that if the result exceed the limits, it should be substituted with the appropriate physical bound. Though, assuming a gate time duration of $t=35$~ns instead of standard $40$~ns, yields results which are more in agreement with the expected theoretical behaviour for decoherence noise, as the deconvoluted results are compatible with one, as expected. This hints that either the quality of the qubit is better then reported in the available calibration data (either due to shorter gate times $t$,
or larger $T_1$ and $T_2$), or that the decoherence model alone poorly describes the noise happening on idle qubit $4$ left interacting with the external environment. However, the good accordance between the deconvoluted results with $t=35$~ns and the experiments suggests the first hypothesis to hold. 

Such conclusion is also corroborated by the experimental results obtained with qubit $25$. In fact, using the deconvolution formulas with reported $T_1$, $T_2$ and standard gate time ($t=40$~ns), we are able to mitigate the effect of noise with good accuracy, as showed in Fig.~\ref{fig:decoherence_noise}(e), hinting that indeed the decay law of the qubit is well described through a decoherence noise model. Also, note that the simulation in Fig.~\ref{fig:decoherence_noise}(c) is tuned with the same noise parameters $p$ and $\gamma$ characterizing qubit $25$. Apart from fluctuations due to, e.g imperfect readout, stochastic measurement outcomes, and noisy Hadamards, there is good agreement between the simulated (red curve in panel (c)) and experimental result (red curve in panel (e)). We do not report analogues experiments using other qubits in the device that produced obviously biased data.

\subsection{Arbitrary Pauli channel}
\begin{figure}[htbp]
    \centering
    \includegraphics[width = 0.5\textwidth]{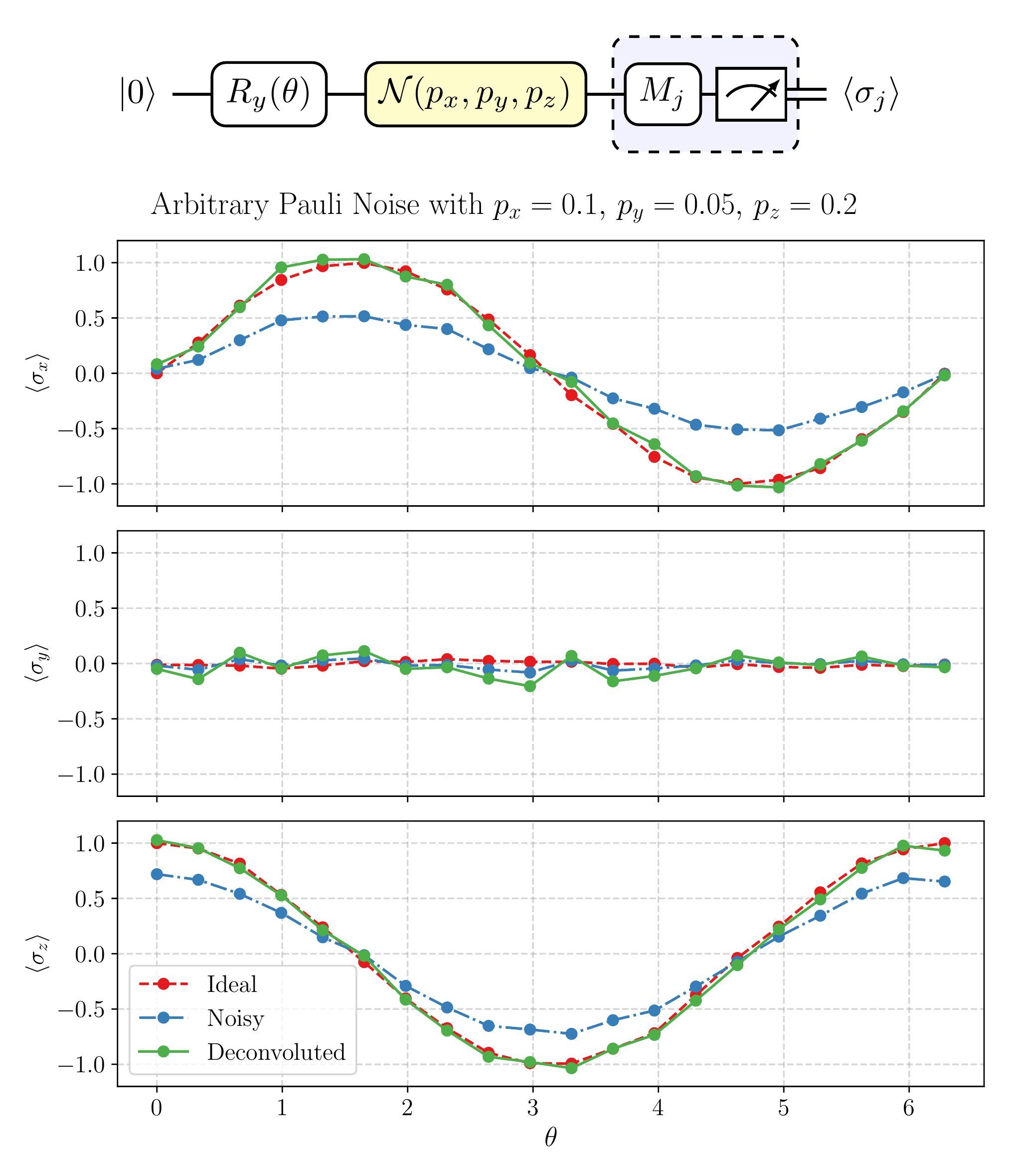}
    \caption{Simulation of the deconvolution process for the general Pauli Channel $\mathcal{N}_{\bm{p}}$~\eqref{eq:general_pauli_channel}. The noise parameters along the three Pauli axes are set to $p_x=0.1, p_y=0.05, p_z=0.2$. The results are obtained simulating the circuit portrayed on top of the image for $n_{\text{shots}} = 1024$ shots and for multiple values of the angle $\theta$. Then, the deconvolution formulas~\eqref{eq:deconvolution_gpc} are used to retrieve the ideal noise-free result. It is clear that the deconvolution effectively mitigates the Pauli noise yielding a final result which is much closer to the ideal noise-free one, up to differences due to stochastic measurement outcomes. In particular, the estimation of $\sigma_y$ is dominated by the statistical error, which is amplified by the correction factor $1/(1-2(p_x+p_z)) = 2.5$.}
    \label{fig:GPC_deconvolution}
\end{figure}

We implemented a simulation of the noise deconvolution of the general Pauli channel~\eqref{eq:general_pauli_channel}, using the quantum virtual machine (QVM) simulator provided with PyQuil~\cite{smith2016practical}. The simulated circuit is showed on top of Figure~\ref{fig:GPC_deconvolution}. A qubit starting in the ground state is rotated in the Bloch sphere around the $y$ axis via $R_y(\theta) = e^{-i\theta\sigma_y/2}$, and then it is subject to the general Pauli noise (yellow box), simulated applying a Pauli transformation chosen randomly with probabilities  $p_x,\,p_y$ and $p_z$. At last, we estimate the expectation value of the three Pauli matrices by appending the appropriate change of basis gate, i.e $M_j \in \{\mathds{1}, H, H S^\dagger\}$ for $\{\sigma_z, \sigma_x, \sigma_y\}$ respectively. 
The noise parameters $(p_x, p_y, p_z)$ are used within the deconvolution formulas~\eqref{eq:deconvolution_gpc} to recover the mitigated results (green curve), which are, as expected, in perfect agreement with the ideal noise-free ones, obtained from executing the quantum circuit without the noisy channel (red curve).

\section{\label{sec:Conclusions} Conclusions}
In conclusion we have shown how mathematically invertible noise maps can always be removed from the final measurement stage, so that one can obtain unbiased expectation values of general observables provided that the noise process in known. We illustrated the method on most known qubit noise maps, and systematically derived their inverse maps (see Table~\ref{tab:tablesummary}). We simulated the noise deconvolution procedure for the case of a general Pauli channel (Fig.~\ref{fig:GPC_deconvolution}) and illustrated our method on noise on actual quantum hardware (Fig.~\ref{fig:decoherence_noise}).   

\section*{\label{sec:Methods} Additional Information}
The experiments with Aspen-9 were performed between the end of October and start of November 2021. The calibration data used in this analysis were downloaded from \url{https://qcs.rigetti.com/lattices} on the 30 October 2021. 

\acknowledgments
This material is based upon work supported by the U.S. Department of Energy, Office of Science, National Quantum Information Science Research Centers, Superconducting Quantum Materials and Systems Center (SQMS) under contract number DE-AC02-07CH11359. We also thank Rigetti (Marco Paini and Matt Reagor in particular) for assistance. S.M. thanks Andrea Mari for a useful discussion.

\bibliography{bibliography}
\newpage

\appendix

\section{\label{app:KarusDecomposition} Kraus Decomposition}
A quantum physical evolution is represented by \textit{(i) linear}, \textit{(ii) completely-positive} and \textit{(iii) trace-preserving} (CPTP) maps taking quantum density operators to quantum density operators. A map satisfying these three properties is called a \textit{quantum channel}, and can be interpreted as a quantum evolution obtained through the interaction of the system with an external environment. A map is a quantum channel if and only if it admits a Kraus (or \textit{operator-sum}) representation as 
\begin{equation}
\label{eq:Kraus_App}
    \rho \longrightarrow \mathcal{E}(\rho) = \sum_{k} A_k \rho A_k^\dagger\,,
\end{equation}
with the trace preserving condition requiring
\begin{equation}
    \Tr \mathcal{E}(\rho) = \Tr \rho \implies \sum_{k} A_k^\dagger A_k = \mathds{1}\, .
\end{equation}
  
The operators $\{A_k\}_k$ are called the Kraus operators of the channel, which are however non-unique~\cite{NielsenChuang}. Such channels are often referred to as \textit{stochastic} channels~\cite{KingMinimalEntropyStates2001, RuskaiAnalysisCompletelypositiveTracepreserving2002}, and if they also preserve the identity ($\mathcal{E}(\mathds{1}) = \mathds{1}$), they are called \textit{unital} (or \textit{bistochastic}). Unitality corresponds to the requirement that also $\sum_{k}A_kA_k^\dagger = \mathds{1}$, from which it is clear that a sufficient condition for a CPTP map to be unital is for its Kraus operators to be self-adjoint $A_k = A_k^\dagger \, \forall k$.  

\section{\label{app:QubitTomographicFormula}Tomographic reconstruction formula for qubits}

In this appendix we show how the tomographic reconstruction formula for systems made of qubits $\mathcal{H} = \mathbb{C}^2$ can be recovered starting from the standard basis expansion in terms of the Pauli matrices~\cite{DMacconePaini2003Spin}. The set of matrices $\{\mathds{1}, \sigma_x, \sigma_y, \sigma_z\}$ form an orthonormal set, and contistutes a basis for the space of $2\times2$ complex matrices $\mathcal{L}(\mathcal{H})=\mathbb{C}^{2\times 2}$. So, given an operator $O \in \mathcal{L}(\mathcal{H})$, the following hold:
\begin{align}
O  & =  \frac{\mathds{1}\Tr[O] + \sigma_x \Tr[O \sigma_x] + \sigma_y \Tr [O \sigma_y] + \sigma_z \Tr[O \sigma_z]}{2}\nonumber\\
& =  \frac{\Tr[O]}{2}\, \mathds{1}+\sum_{\alpha=x,y,z}\frac{\Tr[O\sigma_\alpha]}{2}\sigma_{\alpha}\nonumber\\ 
& =  \sum_{\alpha = x,y,z}\frac{1}{3}\left( \frac{3\Tr[O\sigma_\alpha]}{2}\sigma_{\alpha} + \frac{\Tr[O]}{2}\, \mathds{1} \right)\nonumber\\
& =  \sum_{\alpha=x,y,z}\frac{1}{3}\, \mathcal{E}[O](\sigma_\alpha)\label{eq:TomographyQubit}\ ,
\end{align}
where 
\begin{equation}
\mathcal{E}[O](\sigma_\alpha) = \left( \frac{3\Tr[O\sigma_\alpha]}{2}\sigma_{\alpha} + \frac{\Tr[O]}{2}\mathds{1} \right)\, 
\end{equation}
is the desired \textit{quantum estimator}, with $\{\sigma_x,\sigma_y,\sigma_z\}$ constituting the \textit{quorum} of observables of the tomographic reconstruction.
Equation~\eqref{eq:TomographyQubit} has the same form of the tomographic reconstruction formula in Eq.~\eqref{eq:TomographicRecontructionFormula}, with substitutions
\begin{align*}
\int_{\Lambda} & \longrightarrow   \sum_\lambda \nonumber\\
d\lambda & \longrightarrow  1/3 \nonumber \\
\lambda & \longrightarrow  \{x,y,z\} \nonumber\\
\{Q_\lambda\} & \longrightarrow \{\sigma_x,\sigma_y,\sigma_z\}\, .
\end{align*}
which account for the fact that we are dealing with a discrete, and not continuous, basis expansion. 

Also, note that~\eqref{eq:TomographyQubit} is not the unique choice for the tomographic formula. In fact, one could use a continuous parametrization of the group $SU(2)$ given by operator $\mathcal{R}(\vec{n},\psi )=e^{i\vec{s}\cdot\vec{n}\psi}$,
where $\vec{s}$ is the spin of the particle ($\vec{s}=\vec{\sigma}/2$ for qubits), $\vec{n}=(\cos\phi\sin\theta,\sin\phi\sin\theta,\cos\phi)$, $\theta \in [0,\pi]$ and $\phi,\, \psi \in [0,2\pi]$~\cite{DMacconePaini2003Spin}.

\section{\label{app:QubitNoiseDeconvolution} Noise deconvolution for qubits}
In this appendix we derive the noise deconvolution formula for qubits. Let $\rho$ be a quantum state, and $\mathcal{N}$ a noise channel admitting an inverse map $\mathcal{N}^{-1}$, and $\hat{\mathcal{N}}^{-1}$ its adjoint map. Then, using Eq.~\eqref{eq:inv_noise} in \eqref{eq:qubit_tomography}, yields

\begin{equation}
\begin{aligned}
\label{eq:DeconvolvedQubits}
\expval{O} & = \sum_\alpha \frac{1}{3}\Tr [\hat{\mathcal{N}}^{-1}(\mathbb{E}(O)[\sigma_\alpha])\, \mathcal{N}(\rho)]\\
& = \sum_\alpha \frac{1}{3}\Tr \bigg[ \bigg(\frac{3}{2}\Tr[O\sigma_\alpha] \hat{\mathcal{N}}^{-1}(\sigma_\alpha) + \\
& \hspace{2.5cm} + \frac{1}{2}\Tr [O]\, \hat{\mathcal{N}}^{-1}(\mathds{1}) \bigg)\mathcal{N}(\rho) \bigg]\\
& = \sum_\alpha \frac{1}{3} \Bigg(\frac{3}{2}\Tr [O\sigma_\alpha]\langle\hat{\mathcal{N}}^{-1}(\sigma_\alpha)\rangle_{\mathcal{N}(\rho)} +\\
& \hspace{2.5cm} + \frac{1}{2}\Tr [O] 
\underbrace{\Tr[ \hat{\mathcal{N}}^{-1}(\mathds{1})\mathcal{N}(\rho)]}_{=\Tr[\mathds{1}\rho]=1} \Bigg)\\ 
& = \frac{1}{2}\Tr [O]+\frac{1}{2}\sum_{\alpha=x,y,z}\Tr[O\sigma_\alpha]\langle \hat{\mathcal{N}}^{-1}(\sigma_\alpha)\rangle_{\mathcal{N}(\rho)}\,  . 
\end{aligned}
\end{equation}

Eq.~\eqref{eq:DeconvolvedQubits} lets us deconvolve the effect of noise by evaluating the expectation value of the noise-inverted Pauli matrices $\sigma_\alpha$ on the noisy state $\mathcal{N}(\rho)$. In particular, note that the formula remains valid whether the noise is \textit{unital} --- that is, $\mathcal{N}(\mathds{1})=\mathcal{N}^{-1}(\mathds{1})=\mathds{1}$ --- or not. In fact, in the second line we can always move the adjoint inverse noise $\hat{\mathcal{N}}^{-1}$ from the identity to the noisy state $\mathcal{N}(\rho)$, thus obtaining $\Tr[\hat{\mathcal{N}}^{-1}(\mathds{1})\mathcal{N}(\rho)]=\Tr[\mathds{1}\, \mathcal{N}^{-1}(\mathcal{N}(\rho))] = \Tr[\rho]=1$.

\section{\label{app:inverse_maps} Inverse maps of Noise channels}
In this appendix we explicitly calculate the inverse map of the noise channels discussed in the main text. 

\subsection{Bit-flip, Phase-flip, and Bit-phase-flip channels}
In the following we focus on the Bit-flip channel, but the calculations are identical for the Phase-flip and Bit-Phase-flip channels. 
The Bit-flip is described by Kraus operators $A_0 = \sqrt{1-p}\mathds{1}$ and $A_{1}$ = $\sqrt{p}\,\sigma_x$ so that its action is given by
\begin{equation}
\label{eq:bit-flipApp}
\mathcal{N}_x(\rho) = (1-p)\rho + p\sigma_x\rho \sigma_x\, .
\end{equation}

The Pauli Transfer Matrix $\Gamma_x$ is defined as
\begin{equation}
\label{eq:ptm_x_app}
    (\Gamma_x)_{ij} = \frac{1}{2}\Tr[\sigma_i\, \mathcal{N}_x(\sigma_j)]\, ,
\end{equation}

By straightforward calculation one obtains
\begin{align*}
    (\Gamma_x)_{11} &= (1-p) + p = 1\\
    (\Gamma_x)_{22} &= (1-p) - p = 1-2p\\
    (\Gamma_x)_{33} &= (1-p) - p = 1-2p\\
    (\Gamma_x)_{ij} &= 0\, ,\quad \text{for}\, i\neq j
\end{align*}
thus yielding
\begin{equation}
\Gamma_x = 
    \begin{bmatrix}
    1 & 0 & 0 & 0 \\
    0 & 1 & 0 & 0 \\
    0 & 0 & (1-2p) & 0 \\
    0 & 0 & 0 & (1-2p) \\
    \end{bmatrix}\, ,
\end{equation}
whose inverse is trivially 
\begin{equation}
\Gamma_x^{-1} = 
    \begin{bmatrix}
    1 & 0 & 0 & 0 \\
    0 & 1 & 0 & 0 \\
    0 & 0 & \frac{1}{(2p-1)} & 0 \\
    0 & 0 & 0 & \frac{1}{(2p-1)} \\
    \end{bmatrix}\, .
\end{equation}

The eigenvectors of such Pauli Transfer Matrix are clearly the Pauli matrices $\{|\mathds{1}\rrangle, |\sigma_x\rrangle, |\sigma_y\rrangle, |\sigma_z\rrangle\}$ with eigenvalues $\bm{\lambda} = (1,\,1,\,1/(1-2p),\,1/(1-2p))$. 

The operator sum representation of $\mathcal{N}_x^{-1}$ can be reconstructed by noticing that the map 
\begin{equation}
\label{eq:general_map2}
    \mathcal{E}(O) = \sum_{j=0}^3 \beta_j \sigma_j O \sigma_j\, .
\end{equation}
has also the Pauli matrices as eigenvectors, but with eigenvalues $\bm{\beta} = (\beta_0, \beta_1, \beta_2, \beta_3)$. Since two maps are equals if they have the same action on a basis, then we can find the operator-sum representation of $\mathcal{N}_x^{-1}$ by finding those $\beta_j$ such that $\bm{\lambda} = \bm{\beta}$. If we can find such mapping, then plugging those value in~\eqref{eq:general_map2}, we revcover the operator sum of the inverse map.

The PTM matrix $\Gamma_{\mathcal{E}}$ of $\mathcal{E}$ amounts to
\begin{align*}
    \Gamma_{\mathcal{E}} = \text{diag}(& \beta_0 + \beta_1 + \beta_2 + \beta_3,\\ 
    & \beta_0 + \beta_1 - \beta_2 - \beta_3,\\
    & \beta_0 - \beta_1 + \beta_2 - \beta_3,\\
    & \beta_0 - \beta_1 - \beta_2 + \beta_3)\, ,
\end{align*}

The equality $\Gamma_x^{-1} = \Gamma_\mathcal{E}$ correspond to the system of equations
\begin{equation}
\label{eq:system}
\begin{cases}
1 = \beta_0 + \beta_1 + \beta_2 + \beta_3\\
1 = \beta_0 + \beta_1 - \beta_2 - \beta_3\\
\frac{1}{1-2p} = \beta_0 - \beta_1 + \beta_2 - \beta_3\\
\frac{1}{1-2p} = \beta_0 - \beta_1 - \beta_2 + \beta_3
\end{cases}
\end{equation}
where the first equation is the trace-preserving condition, dictated by the fact that the direct map is TP, and so the inverse map has to be. This condition is also evident from the expression of $\Gamma_x^{-1}$ and $\Gamma_\mathcal{E}$, since the first row is of the form $(1,0,0,0)$. The system of equations~\eqref{eq:system} has solutions
\begin{align*}
    \beta_0 &= \frac{1-p}{1-2p}\\
    \beta_1 &= -\frac{p}{1-2p}\\
    \beta_2 &= \beta_3 = 0
\end{align*}
and substituting these values in Eq.~\eqref{eq:general_map2} leads to the desired operator-sum representation
\begin{equation}
    \mathcal{N}_x^{-1}(O) = \frac{1-p}{1-2p}O - \frac{p}{1-2p}\sigma_x O \sigma_x\, .
\end{equation}

Similarly, the same procedure can be carried out for the Dephasing (generated by $\sigma_z$) and Bit-Phase-flip channel (generated by $\sigma_y$), leading to
\begin{align}
    \mathcal{N}_z^{-1}(O) & = \frac{1-p}{1-2p}O - \frac{p}{1-2p}\sigma_z O \sigma_z\\
    \mathcal{N}_y^{-1}(O) & = \frac{1-p}{1-2p}O - \frac{p}{1-2p}\sigma_y O \sigma_y\, .
\end{align}
Note that for all these three cases the adjoint channels are equal to the direct ones, i.e. $\hat{\mathcal{N}}^{-1} = \mathcal{N}^{-1}$, since the generating operators are all Hermitian (see Appendix~\ref{app:AmplitudeDamping} for a case where this is not true).

We now proceed evaluating the explicit form of the deconvolution formula. Let $\beta \in \{x,y,z\}$ index one of the noise channels $\mathcal{N}_\beta \in \{\mathcal{N}_x, \mathcal{N}_y, \mathcal{N}_z\}$, the action of the inverse map on a Pauli matrix $\sigma_\alpha$ amounts to
\begin{align*}
\mathcal{N}^{-1}_\beta(\sigma_\alpha) & = \frac{1}{1-2p}\bigg((1-p)\, \sigma_\alpha -p\, \sigma_\beta\sigma_\alpha\sigma_\beta \bigg)\\ 
& = \frac{1-2\delta_{\alpha \beta}\,p}{1-2p}\sigma_\alpha\,,
\end{align*}
where in the second line we made use of the fact that $\sigma_\beta\sigma_\alpha\sigma_\beta = (2\delta_{\alpha\beta}-1)\sigma_\alpha$. Substituting this in Eq~\eqref{eq:DeconvolvedQubits}, one obtains
\begin{align*}
\label{eq:deconv_flips}
\langle O \rangle_\beta & = \frac{1}{2}\Tr [O]+\frac{1}{2(1-2p)}\times\\
& \quad \times\sum_{\alpha=x,y,z}\Tr[O\sigma_\alpha]\big(1-2\delta_{\alpha \beta}\,p\big)\expval{\sigma_\alpha}_{\mathcal{N}_\beta(\rho)}\, ,
\end{align*}
where the subscript $\beta$ in $\expval{O}_{\beta}$ is just used to denote that we are deconvolving with respect to noise $\mathcal{N}_\beta$, but remember that it correspond to the mitigated noise-free result. 

Clearly, when the observable to be measured is itself a Pauli matrix $O=\sigma_\gamma$, this further simplifies to
\begin{align*}
\expval{\sigma_\gamma}_\beta  & = \frac{1}{2(1-2p)}\times\\
& \quad \times \sum_{\alpha=x,y,z}\underbrace{\Tr[\sigma_\gamma\sigma_\alpha]}_{=2\delta_{\gamma\alpha}}\big(1-2\delta_{\alpha \beta}\,p\big)\expval{\sigma_\alpha}_{\mathcal{N}_\beta(\rho)}\\ 
& = \frac{1-2\delta_{\gamma\beta}\,p}{1-2p}\expval{\sigma_\gamma}_{\mathcal{N}_\beta(\rho)}\, .
\end{align*}

\subsection{Depolarizing channel}
The Depolarizing channel is represented by the map
\begin{equation*}
    \mathcal{N}_{\text{dep}}(\rho) = \bigg(1-\frac{3p}{4}\bigg)\rho + \frac{p}{4}\bigg(\sigma_x \rho \sigma_x + \sigma_y \rho \sigma_y + \sigma_z \rho \sigma_z\bigg)\, ,
\end{equation*}
having Kraus operators $\{A_0 = \sqrt{1-3p/4}\,\mathds{1},\, A_1 = \sqrt{p}\,\sigma_x/2,\, A_2 = \sqrt{p}\,\sigma_y/2,\, A_3 = \sqrt{p}\,\sigma_z/2\}$. 

By straightforward calculation, the Pauli Transfer Matrix amounts to
\begin{align}
    \Gamma_{\text{dep}} &= 
    \begin{bmatrix}
    1 & 0 & 0 & 0 \\
    0 & 1-p & 0 & 0 \\
    0 & 0 & 1-p & 0 \\
    0 & 0 & 0 & 1-p \\
    \end{bmatrix}\, ,
    \\
    \text{with inverse}\nonumber
    \\
    \Gamma^{-1}_{\text{dep}} &= 
    \begin{bmatrix}
    1 & 0 & 0 & 0 \\
    0 & \frac{1}{1-p} & 0 & 0 \\
    0 & 0 & \frac{1}{1-p} & 0 \\
    0 & 0 & 0 & \frac{1}{1-p} \\
    \end{bmatrix}\, ,
\end{align}

Following the same procedure used for the Bit-flip channel, one arrives at the system of equations
\begin{equation}
    \begin{cases}
    1 = \beta_0 + \beta_1 + \beta_2 + \beta_3\\
    \frac{1}{1-p} = \beta_0 + \beta_1 - \beta_2 - \beta_3\\
    \frac{1}{1-p} = \beta_0 - \beta_1 + \beta_2 - \beta_3\\
    \frac{1}{1-p} = \beta_0 - \beta_1 - \beta_2 + \beta_3
    \end{cases}
\end{equation}
which has solutions $\beta_0 = (4-p)/4(1-p)$ and $\beta_1=\beta_2=\beta_3 = -p/4(1-p)$. Substituting these values in~\eqref{eq:general_map2}, and using the relation $2\Tr[O]\,\mathds{1} = O + \sigma_x O \sigma_x + \sigma_y O \sigma_y + \sigma_z O \sigma_z$, one obtains 
\begin{equation}
    \mathcal{N}_{\text{depol}}^{-1}(O) = \frac{1}{1-p}\left( O-\frac{p}{2}\Tr[O]\mathds{1} \right)\, .
\end{equation}

Plugging this in the tomographic deconvolution formula~\eqref{eq:DeconvolvedQubits}, leads to:
\begin{equation}
    \expval{O} = \frac{1}{2}\Tr[O] + \frac{1}{2}\sum_{\alpha}\frac{\Tr[O\sigma_\alpha]}{1-p}\expval{\sigma_\alpha}_{\mathcal{N}_{\text{dep}}(\rho)}\, ,
\end{equation}
from which is clear that whenever a Pauli matrix is to be measured, $O=\sigma_k$, then the expectation values are contracted by a factor $1-p$, i.e. $\expval{\sigma_k} = \expval{\sigma_k}_\text{dep}/(1-p)$.

\subsection{General Pauli Channel}
The most general channel involving only Pauli operators is the arbitrary Pauli Channel, given by 
\begin{equation}
\mathcal{N}_{\bm{p}}(\rho) = p_0 O + p_x \xox{\rho} + p_y \yoy{\rho} + p_z \zoz{\rho}\, 
\end{equation}
characterized by probabilities $\boldsymbol{p}=(p_0, p_x, p_y, p_z)$, with the trace-preserving condition implying $p_0 = 1-p_x-p_y-p_z$. 
The PTM of this map is diagonal
\begin{align}
    \Gamma_{\bm{p}} = \text{diag}(& 1, p_0 + p_x - p_y -p_z \nonumber\\
    & p_0 - p_x + p_y - p_z, \\
    & p_0 - p_x - p_y + p_z)\, ,\nonumber
\end{align}
and has trivial inverse 
\begin{align}
    \Gamma_{\bm{p}}^{-1} = 
   \text{diag}(& 1, (p_0 + p_x - p_y -p_z)^{-1}\nonumber \\
    & (p_0 - p_x + p_y - p_z)^{-1}, \\
    & (p_0 - p_x - p_y + p_z)^{-1})\nonumber\, .
\end{align}

Again, using the same procedure as before, one arrives at the system of equations: 
\begin{equation}
    \begin{cases}
    1 = \beta_0 + \beta_1 + \beta_2 + \beta_3\\
    \frac{1}{p_0 + p_x -p_y -p_z} = \beta_0 + \beta_1 - \beta_2 - \beta_3\\
    \frac{1}{p_0 - p_x +p_y -p_z} = \beta_0 - \beta_1 + \beta_2 - \beta_3\\
    \frac{1}{p_0 - p_x -p_y +p_z} = \beta_0 - \beta_1 - \beta_2 + \beta_3
    \end{cases}\, ,
\end{equation}
whose solution is reported in Eq.~\eqref{eq:inv_paulichannel} in the main text.

The action of the inverse map on the Pauli matrices is
\begin{align*}
   \mathcal{N}_{\bm{p}}^{-1}(\sigma_x) & = \beta_0\sigma_x+\beta_1\xox{\sigma_x}+\beta_2\yoy{\sigma_x}+\beta_3\zoz{\sigma_x} \\
   & = (\beta_0 + \beta_1 - \beta_2 - \beta_3)\sigma_x \\
   & = \frac{1}{1-2(p_y+p_z)}\sigma_x\,
\end{align*}
and similarly $\sigma_y$ and $\sigma_z$, from which we can have the deconvolution formulas~\eqref{eq:deconvolution_gpc}.

\section{\label{app:AmplitudeDamping} Amplitude Damping}
Amplitude Damping channel is given by the map
\begin{equation}
\begin{aligned}
    & \mathcal{N}_{\text{AD}}(\rho) = K_0\rho K_0^\dagger + K_1\rho K_1^\dagger\, ,\\
    & K_0 = \begin{bmatrix}
    1 & 0 \\ 
    0 & \sqrt{1-\gamma}
    \end{bmatrix}\quad
    K_1 = \begin{bmatrix}
    0 & \sqrt{\gamma}\\
    0 & 0
    \end{bmatrix}\, .
\end{aligned}
\end{equation}

Differently from all the other cases treated above, this channel is not generated by coupled sigma matrices, and in addition one of its generators is not Hermitian. This has two consequences: first, we cannot straightforwardly apply the same eigenvalue matching procedure used above, second one must consider the adjoint channel when deconvolving.

The PTM of Amplitude Damping is
\begin{equation}
    \Gamma_{\text{AD}} = 
    \begin{bmatrix}
    1 & 0 & 0 & 0 \\
    0 & \sqrt{1-p} & 0 & 0 \\
    0 & 0 & \sqrt{1-p} & 0 \\
    p & 0 & 0 & 1-p
    \end{bmatrix}
\end{equation}
whose inverse is
\begin{equation}
    \Gamma_{\text{AD}}^{-1} = 
    \begin{bmatrix}
    1 & 0 & 0 & 0 \\
    0 & \frac{1}{\sqrt{1-p}} & 0 & 0 \\
    0 & 0 & \frac{1}{\sqrt{1-p}} & 0 \\
    \frac{-p}{1-p} & 0 & 0 & \frac{1}{1-p}
    \end{bmatrix}
\end{equation}

In this case the eigenvalues of $\Gamma_{\text{AD}}$ and $\Gamma_{\text{AD}}^{-1}$ are not the Pauli matrices, and so we cannot use the eigenvalue matching with the general map in~\eqref{eq:general_map}. However, the two PTMs have the same structure, so one may easily guess that the operator-sum representation of the two maps share the same operators, something that also always happened in all previous cases. 
Let us then suppose that the inverse map $\mathcal{N}_{\text{AD}}^{-1}$ has the form 
\begin{equation}
\label{eq:ad_mid_app}
    \mathcal{N}_{\text{AD}}^{-1}(\cdot) = \tilde{K}_0 \cdot \tilde{K}_0^{\dagger} - \tilde{K}_1 \cdot \tilde{K}_1^{\dagger}  
\end{equation}
with $\tilde{K}_0 = \dyad{0}+\kappa\dyad{1}$, and $\tilde{K_1} = \tau\ketbra{0}{1}$, with $\kappa\,,\tau$ free parameters to be determined. 
This map has PTM
\begin{equation}
    \Gamma(\kappa,\tau) = 
    \begin{bmatrix}
    \frac{1+\kappa^2-\tau^2}{2} & 0 & 0 & 0 \\
    0 & \kappa & 0 & 0 \\
    0 & 0 & \kappa & 0 \\
    \frac{1-\tau^2-\kappa^2}{2} & 0 & 0 & \frac{1+\tau^2+\kappa^2}{2}
    \end{bmatrix}\, ,
\end{equation}

and by requiring that $\Gamma(\kappa,\tau)=\Gamma_{\text{AD}}^{-1}$, we obtain
$$
\kappa = \frac{1}{\sqrt{1-\gamma}}\,,\quad \tau = \sqrt{\frac{\gamma}{1-\gamma}}\, , 
$$
thus recovering the inverse map
\begin{equation}
\begin{aligned}
& \mathcal{N}_{\text{AD}}^{-1}(O) = \tilde{K}_0 O \tilde{K}_0^\dagger - \tilde{K}_1 O \tilde{K}_1^\dagger\, ,\\
& \tilde{K}_0 = 
\begin{bmatrix}
1 & 0 \\
0 & \frac{1}{\sqrt{1-\gamma}}
\end{bmatrix}
\,,\,
\tilde{K}_1 = 
\begin{bmatrix}
0 & \sqrt{\frac{\gamma}{1-\gamma}} \\ 
0 & 0
\end{bmatrix}\, .
\end{aligned}
\end{equation}

In order to evaluate the deconvolution formula, we first need to calculate the adjoint of the inverse channel. Be $\Phi$ a linear map, its adjoint $\hat{\Phi}$ is defined as the unique map satisfying the following relation
\begin{equation}
\langle A, \Phi(B)\rangle = \langle \hat{\Phi}(A), B \rangle\, . 
\end{equation}
where $\langle \cdot, \cdot \rangle$ denotes the Hilbert-Schmidt inner product $\langle A, B\rangle \equiv \Tr[A^\dagger B]$.
Let's consider a generic linear map of the form 
\begin{equation}
\label{eq:general_map_for_adjoint}
\Phi(A) = \sum_{k}\alpha_k\, V_k A V^\dagger_k\, , \quad  \alpha_k \in \mathbb{R}\, .
\end{equation}
which is not, in general, nor CP or TP, since we make no further hypothesis on $\alpha_k$ and $V_k$. By direct application of the definition of adjoint map, we obtain
\begin{align*}
\langle A, \Phi(B)\rangle & \equiv \Tr[A^\dagger \Phi(B)] = \Tr\left[ A^\dagger \sum_k \alpha_k\, V_k B V_k^\dagger \right] \\
& = \Tr[\sum_k \alpha_k\,  V_k^\dagger A^\dagger V_k\, B]\\
& = \Tr \left[ \left(\sum_k \alpha_k V_k^\dagger A V_k\right)^\dagger B\right]\\
& = \left\langle \sum_k \alpha_k V_k^\dagger A V_k, B \right\rangle \\
& \quad \Rightarrow \hat{\Phi}(A) = \sum_k \alpha_k V_k^\dagger A V_k\, ,
\end{align*}
where we used the linearity and cyclic property of the trace, as well as the fact that the coefficients are real, $\alpha_k^*=\alpha_k \in \mathbb{R}$. We see that for any map of the form~\eqref{eq:general_map_for_adjoint}, its adjoint is obtained by simply substituting the operators with their adjoint, i.e. $V_k \rightarrow V_k^\dagger$. If the map $\Phi$ leverages only hermitian operators $V_k=V_k^\dagger$, as it happens with every Pauli noise channel, than the adjoint and the direct map of course coincides, $\hat{\Phi}(\cdot) = \Phi(\cdot)$. However, the Amplitude Channel uses non Hermitian generators $V_k$, thus has a non-trivial, yet simple, adjoint map.

Straightforward application of the deconvolution formula then leads to the deconvolved expectation values
\begin{equation}
\begin{aligned}
\expval{\sigma_x} & = \frac{1}{\sqrt{1-\gamma}}\langle \sigma_x \rangle_{\mathcal{N}_{\text{AD}}(\rho)} \\
\expval{\sigma_y} & = \frac{1}{\sqrt{1-\gamma}}\langle \sigma_y \rangle_{\mathcal{N}_{\text{AD}}(\rho)} \\
\expval{\sigma_z} & = \frac{1}{1-\gamma}\qty(\langle \sigma_z \rangle_{\mathcal{N}_{\text{AD}}(\rho)} -\gamma)\, .
\end{aligned}
\end{equation}

\section{\label{app:TwoKraus} 2-Kraus channel}
The set of channels considered here is generated by two parametrized Kraus operators
\begin{equation}
\mathcal{N}_{\text{two}}(\rho) = \sum_{i=1, 2}A_i\rho A_i^\dagger\, ,  
\end{equation}
with $A_1 = \cos\alpha\dyad{0}+\cos\beta\dyad{1}$, and $A_2 = \sin\beta\ketbra{0}{1}+\sin\alpha\ketbra{1}{0}$. The PTM of this channel is
\begin{widetext}
\begin{equation}
\Gamma_{\text{two}} =
\begin{bmatrix}
1 & 0 & 0 & 0 \\
0 & \cos(\alpha-\beta) & 0 & 0 \\
0 & 0 & \cos(\alpha+\beta) & 0 \\
\frac{\cos(2\alpha)-\cos(2\beta)}{2} & 0 & 0 & \frac{\cos(2\alpha)+\cos(2\beta)}{2}
\end{bmatrix}
\end{equation}
with inverse
\begin{equation}
\Gamma_{\text{two}}^{-1} =
\begin{bmatrix}
1 & 0 & 0 & 0 \\
0 & \frac{1}{\cos(\alpha-\beta)} & 0 & 0 \\
0 & 0 & \frac{1}{\cos(\alpha+\beta)} & 0 \\
\frac{\cos(2\beta)-\cos(2\alpha)}{\cos(2\alpha)+\cos(2\beta)} & 0 & 0 & \frac{2}{\cos(2\alpha)+\cos(2\beta)}
\end{bmatrix}\, .
\end{equation}
\end{widetext}

Using the trigonometric relation 
\begin{align*}
    \cos(2\alpha)+\cos(2\beta) &= 2\cos(\frac{2\alpha-2\beta}{2})\cos(\frac{2\alpha+2\beta}{2})\\
    &= 2\cos(\alpha-\beta)\cos(\alpha+\beta)
\end{align*}
we can rewrite the elements of $\Gamma_{\text{two}}^{-1}$ as
\begin{align*}
    (\Gamma_{\text{two}}^{-1})_{11} &= h_{\alpha\beta} \cos(\alpha+\beta)\\
    (\Gamma_{\text{two}}^{-1})_{22} &= h_{\alpha\beta} \cos(\alpha-\beta)\\
    (\Gamma_{\text{two}}^{-1})_{33} &= h^2_{\alpha\beta} \frac{\cos(2\alpha)+\cos(2\beta)}{2}\\
    (\Gamma_{\text{two}}^{-1})_{30} &= h_{\alpha\beta}\frac{\cos(2\beta)-\cos(2\alpha)}{2}\, ,
\end{align*}
with $h_{\alpha\beta} = \frac{2}{\cos(2\alpha)+\cos(2\beta)}$. Written in this way, these matrix elements are very similar to those in the PTM of the direct channel $\Gamma_{\text{two}}$. The differences are in the presence of the pre-factor $h_{\alpha\beta}$, as well as in the signs of the angles in elements `$11$' and `$22$', and in the sign in the difference in element `$30$'.
This suggest that the operator-sum representation of the inverse map can be obtained starting from the direct one with some small changes, as it happened with the amplitude damping channel. First of all, we can multiply the Kraus operators by $\sqrt{h_{\alpha\beta}}$ to introduce the pre-factor, then, to account for the difference in elements `$11$' and `$22$', we can subtract the two operators instead of summing them. At last, element `$30$' can be fixed by changing $\alpha \leftrightarrow \beta$ in the first Kraus operator $A_1$.  Incidentally, these changes also fix the `$44$' element to the correct value. Eventually, making these changes leads to defining new operators
\begin{align*}
    & B_1  = \sqrt{h_{\alpha\beta}}\cos(\beta)\dyad{0} + \sqrt{h_{\alpha,\beta}}\cos(\alpha)\dyad{1}\\
    & B_2  = \sqrt{h_{\alpha\beta}}\sin(\beta)\ketbra{0}{1}+\sqrt{h_{\alpha,\beta}}\sin(\alpha)\ketbra{1}{0}\\
    & h_{\alpha\beta} = \frac{2}{\cos(2\alpha)+\cos(2\beta)}\, ,
\end{align*}
to be used within the inverse map 
$$
\mathcal{N}_{\text{two}}^{-1}(\cdot) = B_1 \cdot B_1^\dagger - B_2 \cdot B_2^\dagger\, .
$$
One can check that this map has the desired Pauli Transfer Matrix $\Gamma_{\text{two}}^{-1}$.

As with the amplitude damping case, one the generators ($B_2$) is not Hermitian, thus one be careful in considering the adjoint inverse map when evaluating the deconvolved mean values. By explicit calculations the following holds:
\begin{align}
\expval{\sigma_x} & = \frac{1}{\cos(\alpha-\beta)}\expval{\sigma_x}_{\mathcal{N}_{\text{two}}(\rho)} \nonumber\\
\expval{\sigma_y} & = \frac{1}{\cos(\alpha+\beta)}\expval{\sigma_y}_{\mathcal{N}_{\text{two}}(\rho)} \\
\expval{\sigma_z} & = h_{\alpha\beta}\big(\cos^2(\beta)+\sin^2(\alpha)-1+\expval{\sigma_z}_{\mathcal{N}_{\text{two}}(\rho)}\big)\,.\nonumber
\end{align}

\end{document}